\documentclass[%
 aip,
 amsmath,amssymb,
 reprint,%
]{revtex4-1}

\usepackage{graphicx}
\usepackage{dcolumn}
\usepackage{bm}

\usepackage[utf8]{inputenc} 
\usepackage[T1]{fontenc} 
\usepackage{mathptmx} 
\usepackage{etoolbox}
\usepackage{braket}    
\usepackage{nicefrac}
\usepackage{rotating}
\usepackage{makecell}
\usepackage{multirow}
\usepackage{array}
\usepackage{siunitx}
\usepackage[version=3]{mhchem} 

\newcommand{\vcf}{\hat{V}^{\mathrm{CF}}}
\newcommand{\hso}{\hat{H}^{\mathrm{SO}}}

\newcommand{\okq}{\hat{O}_k^q}
\newcommand{\bkq}{B_k^q}
\newcommand{\jmj}{ \ket{j,m_j}}
\newcommand{\lml}{ \ket{l,m_l}}
\newcommand{\lmlsms}{ \ket{l,m_l,s,m_s}}
\newcommand{\JMJ}{ \ket{J,M_J}}
\newcommand{\LML}{ \ket{L,M_L}}
\newcommand{\icm}{\mathrm{cm}^{-1}}

\makeatletter
\def\@email#1#2{%
 \endgroup
 \patchcmd{\titleblock@produce}
  {\frontmatter@RRAPformat}
  {\frontmatter@RRAPformat{\produce@RRAP{*#1\href{mailto:#2}{#2}}}\frontmatter@RRAPformat}
  {}{}
}%
\makeatother
\begin{document}

\preprint{AIP/123-QED}

\title{\emph{Ab initio} derivation of the crystal field parameters for lanthanide ions: The f$^1$ case}

\author{Dumitru-Claudiu Sergentu}
\affiliation{Faculty of Chemistry, ''Alexandru Ioan Cuza'' University of Ia\textcommabelow{s}i, Ia\textcommabelow{s}i, Romania} 
\affiliation{ICI-RECENT AIR (RA03), ''Alexandru Ioan Cuza'' University of Ia\textcommabelow{s}i, Ia\textcommabelow{s}i, Romania}

\author{Gwenhaël Duplaix-Rata}
\affiliation{Univ Rennes, CNRS, ISCR (Institut des Sciences Chimiques de Rennes) – UMR
6226, F-35000 Rennes, France}

\author{Ionel Humelnicu}
\affiliation{Faculty of Chemistry, ''Alexandru Ioan Cuza'' University of Ia\textcommabelow{s}i, Ia\textcommabelow{s}i, Romania}

\author{Boris Le Guennic}
\affiliation{Univ Rennes, CNRS, ISCR (Institut des Sciences Chimiques de Rennes) – UMR
6226, F-35000 Rennes, France}

\author{R\'emi Maurice}
\affiliation{Univ Rennes, CNRS, ISCR (Institut des Sciences Chimiques de Rennes) – UMR
6226, F-35000 Rennes, France}

\email{remi.maurice@univ-rennes.fr}
\email{dumitru.sergentu@uaic.ro}

\date{\today}

\begin{abstract}
The crystal field theory as explained by Abragam and Bleaney in their landmark 1970 book on transition-ion electron paramagnetic resonance remains a cornerstone in the development of luminescence applications and molecular magnets based on the $f$-elements. The modern numerical derivation of the 27 $B_k^q$ Stevens crystal field parameters (CFPs), which describe the splitting of the energy levels of a central ion, is traditionally achieved through the effective Hamiltonian theory and multiconfiguration wavefunction theory calculations, insofar as the lowest $J$ level fully captures the targeted low-energy physics. In this work, we present a novel theoretical approach for determining the CFPs. The procedure resembles the traditional extraction path but crucially accounts for the full $\ket{J,M_J}$ space of an ion configuration with $L=3$ and $S=\nicefrac{1}{2}$. By demonstrating the extraction procedure using the simplest case of a Ce$^\text{III}$ 4f$^1$ ion with a crystal-field split $J \in \{\nicefrac{5}{2}, \nicefrac{7}{2}\}$ manifold, it is shown for the first time that a unique set of CFPs describes the splitting and mixing both the $J$ manifolds. In fact, this $J/J^\prime$ mixing is analogous to the ``spin mixing'' in binuclear transition metal complexes. At the employed level of calculation, we demonstrate that there is no spin-orbit coupling influence on the CFP values, contrary to previous beliefs. Moreover, for the 4f$^1$ case, the present extraction yields crystal-field and spin–orbit parameters similar to those obtained from \emph{ab initio} ligand field theory (AILFT). This work represents the first step of a larger effort in reviewing the theory and extraction procedures of CFPs in f-element complexes.
\end{abstract}

\maketitle

\section{\label{sec:introduction}Introduction\protect\\ }

Since its conceptualization, the crystal field (CF) theory has served as a foundational framework for understanding traits of the magnetic susceptibility, EPR and other spectroscopy properties of organometallic complexes.\cite{Abragam:1970a} Over the years, the theory has evolved to provide increasingly accurate descriptions of the electronic structure of transition metal and f-block ions, with direct implications in modern luminescence,\cite{Natrajan:2012a, Tessitore:2023a} magnetism,\cite{Rinehart:2011a, Liddle:2015a, Coronado:2020a, Bernot:2023a} and quantum technologies.\cite{Mcadams:2017a, Moreno:2025} Focus is directed to 4f and 5f ions since these are essential components for single-molecule magnets (SMMs)\cite{Woodruff:2013a, Layfield:2015a, Escalera:2019a, Dey:2025a, Pointillart:2017a} due to the diversity of their orbital structures.\cite{Kaltsoyannis:2024a} This has spurred interdisciplinary efforts to explore their potential for memory effects that operate closer to room temperature,\cite{Guo:2017a, Guo:2018a, Guo:2022a, Gould:2022a, Goodwin:2017a} particularly following the discovery of the first 4f-based SMM, [Tb(Pc)$_2$]$^-$, in the early 2000's.\cite{Ishikawa:2003a, Ishikawa:2004a}

The method traditionally employed for determining CF parameters (CFPs), \emph{i.e.} the phenomenological parameters that effectively model the impact of the ligand coordination on a central ion’s energy states,\cite{Duan:2010a} relies on experimental data fitted with an underlying electrostatic treatment of the metal-ligand interaction. While the approach proves effective, easily realized through softwares such as PHI,\cite{Chilton:2013a} CONDON,\cite{Schilder:2004a} SIMPRE,\cite{Baldovi:2013a, Baldovi:2014a} MAGPACK,\cite{Borras:2001a} or PyCrystalField,\cite{Scheie:2021a} it remains reliant on high symmetry CFs and introduction of (empirical) orbital-reduction parameters to account for covalent bonding. Apart from these popular choices, NJA-CFS is a most recent and versatile software toolkit,\cite{Fiorucci:2015a} allowing the non-experienced user to choose among different routes to derive CFPs. Nowadays, the extraction of CFPs shifted to solving model (analytical) Hamiltonians through one-to-one correspondence with (effective) numerical Hamiltonians derived from \emph{ab initio} wavefunction calculations.\cite{Chibotaru:2008a, Chibotaru:2012a, Gendron:2014a, Gendron:2015a, Ungur:2017a, Riccardo:2018a, Bolvin:2023a, Islam:2023a} This is typically achieved using complete active space self-consistent field (CASSCF) calculations,\cite{Roos:1980a, Roos:1980b} which account for the multiconfigurational nature of the electronic states in f-complexes, enable systematic improvement of the state energies by (multireference) second-order perturbation theory (PT2),\cite{Andersson:1990a, Andersson:1992a, Finley:1998a, Angeli:2001a} and allow an \emph{a posteriori} treatment of the spin-orbit coupling (SOC) by spin-orbit configuration interaction (SOCI).\cite{Malmqvist:2002a} The \texttt{SINGLE\_ANISO} package,\cite{Chibotaru:2012a} available in well-known electronic structure software such as ORCA\cite{Neese:2012a, Neese:2022a} and OpenMolcas,\cite{Manni:2023a} is nowadays a popular choice for the \emph{ab initio} extraction of CFPs\cite{Gould:2022a, Zhu:2024a, Islam:2023a, Kushvaha:2024a, Corner:2025a} through the so-called pseudospin and irreducible tensor operator (ITO) techniques.\cite{Chibotaru:2012a, Ungur:2017a} The ab initio ligand field theory (AILFT), introduced by Atanasov and coworkers\cite{Atanasov:2015a}, is an alternative approach for extracting CFPs from first principles. Comparing with the previous route, the parameters are now extracted directly at the orbital level rather than from the coupled many-electron states, and the procedure involves a final step of fitting to many-electron state energies instead of using an ITO decomposition.\cite{Aravena:2016a, Jung:2017a, Jung:2019a, Lang:2020a, Islam:2023a, Autillo:2024a}

The mathematical formalism\cite{Stevens:1952a, AbragamBleaney, Rudowicz:2004a, Duros:2025a} underpinning the determination of CFPs utilizes the CF potential, $\vcf$, constructed from Stevens equivalent operators $\okq$:
\begin{equation}
    \hat{V}^{\text{CF}}(X) = \sum_{k=2,4,6} a_k^X \sum_{q=-k}^{k} B_k^q \hat{O}_k^q(X)
    \label{Eq:1}
\end{equation}
\noindent Here, \( k \in \{2, 4, 6\} \) denotes the \textit{rank} of the operator, restricted to even integers due to time-reversal and spatial symmetry constraints. The index \( q \in [-k, k] \) represents the \textit{order} of the operator, analogous to the magnetic quantum number in spherical harmonics, and defines the orientation-dependent component within each rank. The $\bkq$ coefficients, hereafter referred as $B_k^q(\text{Stevens})$, are the Stevens CFPs to be determined. Within the point-charge electrostatic model, $B_k^q(\text{Stevens}) = A_k^q \langle r^k \rangle$, where the coefficients $A_k^q$ represents the geometric lattice sums and $\langle r^k \rangle$ are radial expectation values. In a given basis-state representations \( X \): namely, the orbital basis $\lml$, the spin-free many-electron basis $\LML$, and the spin–orbit coupled basis $\JMJ$ (or $\jmj$ for one-electron cases), \( a_k^X \) (Stevens coefficients) are reduced matrix elements specific to the magnetic ion and $X$ manifold under consideration ($l$, $L$ or $J$).\cite{AbragamBleaney, Riccardo:2018a} These are tabulated\cite{Abragam:1970a} if not calculated on-the-fly, and are introduced to ensure consistency of the $\bkq$ values across the different basis-state representations \( X \).

For a given f-ion configuration, the lowest-energy $J$-manifold is commonly chosen as the model basis set for expanding the analytical matrix of \( \hat{V}^{\mathrm{CF}} \) and construction of the \emph{ab initio} numerical Hamiltonian from spin-orbit coupled wavefunctions, $\hso$. The latter is derived by expressing the diagonal matrix of SOCI energies in the eigenbasis of the Zeeman Hamiltonian, itself being constructed in the space of the lowest \( 2J+1 \) SOCI wavefunctions~\cite{Ungur:2017a}. $\hso$ is subsequently mapped onto the analytical matrix of $\vcf$, allowing for the extraction of the complete set of 27 CFPs via the ITO decomposition.\cite{Chibotaru:2012a, Ungur:2017a} An additional set of $\bkq$ parameters is derived in the spin-free orbital basis \( \ket{L, M_L} \). Few studies acknowledge that the CFPs determined in the $J$ vs.\ a spin-free basis are somewhat similar, although not identical. These include the \ce{[Ln(DPA)3]^{3-}} series,\cite{Jung:2019a} \ce{PrCl3} and few lanthanide sandwich complexes,\cite{Riccardo:2018a} and a recently reported Yb(III) alkyl complex.\cite{Ashuiev:2024a} In other reports, involving actinides in particular, the CFPs extracted before and after spin--orbit coupling may differ significantly,\cite{Ashraful:2023a, Autillo:2020a, Autillo:2024a} acknowledging that the parameters obtained after the introduction of the SOC are irregular, likely as a consequence of significant $J$-mixing. Certainly, there is need for a clarification on the actual role of the SOC and of the $J$-mixing on the spectra (energies and wavefunctions) and on the extraction of the CFPs.  In fact, as Bolvin et al.\ suggested, the CFPs extracted for a given $J$-manifold via the ITO procedure are representative only of the splitting within that specific manifold and are thus relevant solely to spectroscopies probing that manifold.\cite{Jung:2019a} In other words, the truncation of the $J$-manifold leads to the differences observed in the parameters extracted before and after SOC, essentially because the potential $J$-mixing induced by the CF interaction (off-diagonal crystal field terms) is generally not permitted in models retaining only the lowest-energy $J$-manifold.

This article resolves the aforementioned issue by introducing a procedure to calculate CFPs for the \ce{f^1} configuration, accounting for SOC, fully \emph{ab initio}, without truncating the $J$-manifold stemming from the same ground orbital term. We show that the model matrix of $\vcf$ can be properly derived in this ``complete'' $J$ basis set by relying only on the Stevens formalism and Clebsch-Gordan (CG) algebra. The model can be extracted with data from widespread CASSCF and SOCI calculations. We show that the CFPs extracted by this procedure are not merely similar but strictly \emph{identical} regardless of the introduction of the SOC, resolving a confusion that flourished over the past decades. To ease the discussion, the extraction procedure is showcased for a \ce{Ce^{3+}} f$^1$ configuration in a field of point charges exerting a CF of $D_\text{4h}$ symmetry. After presenting CFPs for various other point-group symmetries, we calculate CF and magnetic data for the anion of cerocene, Ce(C$_8$H$_8$)$_2^-$. In this realistic case, the procedure correctly models the CF admixture of $J=7/2$ excited levels into the ground state (GS) $J=5/2$ Kramers doublet, therefore explaining the anisotropy of its characteristic (effective) $g$-tensor.\cite{Walter:2009a}

\section{\label{sec:computational-details}Computational Details\protect\\ }

CASSCF\cite{Roos:1980a, Roos:1980b} calculations for the seven spin-doublet roots of the $f^1$ configuration were performed with ORCA (v6.1)\cite{Neese:2025a} for the cerocene anion, Ce(C$_8$H$_8$)$_2^-$, and a series of models with a \ce{Ce^3+} ion in a field of $q=-2$ point charges at 3 \AA{} exerting a CF of particular symmetry: $D_{\infty\text{h}}$ (linear) with two charges along the $z$-axis, $D_\text{4h}$ with 4 charges in a square-planar geometry in the $xy$-plane, $O_\text{h}$ with six charges in an octahedron geometry, $C_\text{2v}$ with two charges in a water configuration, $C_\text{3v}$ with 3 charges in an ammonia configuration, and finally $T_\text{d}$ with 4 charges in a methane configuration. In these models, the cerium ion was treated with the all-electron SARC2-DKH-QZVP basis set.\cite{Aravena:2016b} Concerning Ce(C$_8$H$_8$)$_2^-$, the experimental structure was retained,\cite{Kilimann:1994a} with a Ce--ring distance of 2.043 \AA{} and averaged Ce--C and C--C bond distances of 2.743 and 1.404 \AA{} respectively. Cerium and the remaining atoms were treated with all-electron SARC2-DKH-QZVP\cite{Pantazis:2009a} and DKH-def2-TZVP basis sets,\cite{Weigend:2005a} respectively. Cartesian coordinates for all the Ce$^{3+}$ point-charge models and the cerocene anion are provided in Tables \ref{tab-si:model_coords} and \ref{tab-si:cerocene_coords} of the Supplementary Material file).

Unless for Ce(C$_8$H$_8$)$_2^-$, all the CAS-level calculations used the resolution of identity approximation (RIJCOSX)\cite{Paris:2021a,Neese:2003a} for the Coulomb and exchange integrals, large automatically-generated auxiliary basis sets by the AutoAux procedure,\cite{Stoychev:2017a} and the second-order Douglas-Kroll-Hess (DKH2) Hamiltonian\cite{Reiher:2004a} to account for scalar relativistic effects. For each complex, the seven spin-doublet CASSCF states and energies were subsequently used in a second step to account for SOC within the SOCI framework.\cite{Neese:2005a} The \emph{ab initio} energies obtained in CASSCF and SOCI calculations are gathered in Tables \ref{tab-si:energies-casscf} and \ref{tab-si:energies-casscf-so}, respectively. 

The theoretical workflow, detailed in this article, constitutes the basis for the first release of the \texttt{NewMag} code, which is publicly available on GitHub at: \\\texttt{https://github.com/clausserg/newmag.git}

\section{\label{sec:cfps-sof}Extraction of CF parameters in the spin-free formalism\protect\\ }

\subsection{\label{subsec:cfps-sof-model}Construction of the model CF matrix of $\vcf$\protect\\ }

In the spin-free formalism, the f$^1$ configuration of a free ion leads to a sevenfold orbital degenerated $^2F$ term ($l=3$ and $s=1/2$), with the individual components labeled by the orbital projection $m_l \in [-l, +l]$. Thus, the analytical matrix of $\vcf$ is constructed by expanding the Stevens equivalent operators $\okq$ of Eq.~\ref{Eq:1} in the model space spanned by the seven orbital basis states $\ket{l=3, m_l}$. To exemplify, for a $D_\text{4h}$ CF described by square-planar coordination in the $xy$-plane and the fourfold rotation axis along the cartesian $z$-axis, only operators with $q=0$ and $q=4$ remain invariant under the point group symmetry operations, and thus contribute non-zero matrix elements in the CF matrix. The expression for the CF operator simplifies in this particular case to:
\begin{align}
\vcf(l) &= a_2^l \left[B_2^0 \hat{O}_2^0(l)\right]
          + a_4^l \left[B_4^0 \hat{O}_4^0(l) + B_4^4 \hat{O}_4^4(l)\right] \nonumber \\
        &\quad + a_6^l \left[B_6^0 \hat{O}_6^0(l) + B_6^4 \hat{O}_6^4(l)\right]
\label{Eq:2}
\end{align}
\noindent where $a_2^l=-2/45$, $a_4^l=+2/495$ and $a_6^l=-4/3861$ are the reduced matrix elements for $k=2$, 4 and 6, respectively, and an $l$ orbital-basis representation.\cite{Abragam:1970a} Consequently, the only CFPs that survive in the CF matrix are the so-called axial terms $B_2^0$, $B_4^0$ and $B_6^0$, and tetragonal terms $B_4^4$ and $B_6^4$, reflecting the axial and fourfold character of the $D_\text{4h}$ interaction. The resulting matrix representation of $\vcf$ is provided in Table \ref{tab:vcf-mod}. The five CFPs, encoding here the energetic splitting of the seven f orbitals in $D_\text{4h}$, are the unknowns that need to be determined by comparing the analytical matrix elements in Table \ref{tab:vcf-mod} with counterparts from a numerical (effective) CF Hamiltonian matrix, $\hat{H}^\text{(eff)}$, which is derived from \emph{ab initio} calculations in the following subsection.

\begin{sidewaystable}
\centering
\caption{Matrix elements of $\vcf$ expressed in the $\ket{l=3, m_l}$ basis set of an f$^1$ configuration in a $D_\text{4h}$ crystal field.} 
\label{tab:vcf-mod}
\renewcommand{\arraystretch}{2}
\begin{tabular}{rccccccc}
\toprule
$\vcf$ & 
$\ket{3, -3}$ & $\ket{3, -2}$ & $\ket{3, -1}$ & $\ket{3, 0}$ & $\ket{3, 1}$ & $\ket{3, 2}$ & $\ket{3, 3}$ \\
\hline
$\bra{3, -3}$ & $- \frac{2 B^{0}_{2}}{3} + \frac{8 B^{0}_{4}}{11} - \frac{80 B^{0}_{6}}{429}$ & $0$ & $0$ & $0$ & $\frac{8 \sqrt{15} B^{4}_{4}}{165} - \frac{80 \sqrt{15} B^{4}_{6}}{1287}$ & $0$ & $0$ \\
$\bra{3, -2}$ & $0$ & $- \frac{56 B^{0}_{4}}{33} + \frac{160 B^{0}_{6}}{143}$ & $0$ & $0$ & $0$ & $\frac{8 B^{4}_{4}}{33} + \frac{160 B^{4}_{6}}{429}$ & $0$ \\
$\bra{3, -1}$ & $0$ & $0$ & $\frac{2 B^{0}_{2}}{5} + \frac{8 B^{0}_{4}}{33} - \frac{400 B^{0}_{6}}{143}$ & $0$ & $0$ & $0$ & $\frac{8 \sqrt{15} B^{4}_{4}}{165} - \frac{80 \sqrt{15} B^{4}_{6}}{1287}$ \\
$\bra{3, 0}$ & $0$ & $0$ & $0$ & $\frac{8 B^{0}_{2}}{15} + \frac{16 B^{0}_{4}}{11} + \frac{1600 B^{0}_{6}}{429}$ & $0$ & $0$ & $0$ \\
$\bra{3, 1}$ & $\frac{8 \sqrt{15} B^{4}_{4}}{165} - \frac{80 \sqrt{15} B^{4}_{6}}{1287}$ & $0$ & $0$ & $0$ & $\frac{2 B^{0}_{2}}{5} + \frac{8 B^{0}_{4}}{33} - \frac{400 B^{0}_{6}}{143}$ & $0$ & $0$ \\
$\bra{3, 2}$ & $0$ & $\frac{8 B^{4}_{4}}{33} + \frac{160 B^{4}_{6}}{429}$ & $0$ & $0$ & $0$ & $- \frac{56 B^{0}_{4}}{33} + \frac{160 B^{0}_{6}}{143}$ & $0$ \\
$\bra{3, 3}$ & $0$ & $0$ & $\frac{8 \sqrt{15} B^{4}_{4}}{165} - \frac{80 \sqrt{15} B^{4}_{6}}{1287}$ & $0$ & $0$ & $0$ & $- \frac{2 B^{0}_{2}}{3} + \frac{8 B^{0}_{4}}{11} - \frac{80 B^{0}_{6}}{429}$ \\ \toprule
\end{tabular}
\end{sidewaystable}

\subsection{\label{subsec:cfps-sof-numerical}Derivation of the CF matrix from \emph{ab initio} calculations\protect\\ }

A state-averaged CASSCF calculation is first performed over the seven spin-doublet CF states of the f$^1$ configuration and the wavefunctions are expressed and collected in the Slater determinant basis. The active space and number of states calculated are not limited to seven metal f orbitals and seven crystal field states, but must include these at a minimum. In case of spurious mixing or hybridization among the active space orbitals, unitary rotations should be performed to localize and easily identify the seven f orbitals with their corresponding $m_l$ quantum numbers. With these wavefunction data, an eigenvector matrix is constructed in the $\ket{l=3,m_l}$ model space, where each row corresponds to a CASSCF state wavefunction $\psi_k$ (i.e.\ contains the coefficients of the Slater determinant expansion of $\psi_k$ in the localized f-orbital basis). For active spaces larger than the seven f orbitals, a L\"owdin symmetric orthonormalization\cite{Lowdin:1950a} would be performed in order to project each individual CASSCF wavefunction onto the $\ket{l=3,m_l}$ model space. The procedure then generates a set of L\"owdin orhtonormalized wavefunctions, $\psi_k^L$. Consequently, in these cases, each row in the CASSCF eigenvector matrix contains the Slater determinant expansion coefficients of $\psi_k^L$.

With the CASSCF wavefunction matrix constructed above, the numerical (effective) Hamiltonian, $\hat{H}^\text{(eff)}$, is formed by means of the des Cloizeaux formalism\cite{Cloizeaux:1960a} in the $\ket{l=3,m_l}$ model space of real spherical harmonics (RSH):
\begin{equation}
\label{eq:cloizeaux}
\hat{H}^\text{(eff),RSH} = \sum_k E_k \braket{\phi_i | \psi_k^{(L)}} \braket{\psi_k^{(L)} | \phi_j}
\end{equation}
\noindent where $\phi_i$, $\phi_j$ are model-space $\ket{l=3, m_l}$ orbital functions, $\psi_k^{(L)}$ are the (L\"owdin orthonormalized) CASSCF wavefunctions in the Slater determinant basis, and $E_k$'s are the corresponding CASSCF energies. These energies may, of course, be replaced by more accurate values that include corrections for dynamic electron correlation, if available. Finally, a basis transformation is performed in order to translate the (effective) Hamiltonian from the RSH basis into the complex spherical harmonics basis (CSH), where $m_l$ is a well defined quantum number. The basis-transformation matrix $U$, provided in Table \ref{tab-si:rsh-csh}, is easily constructed according to Blanco et al.\cite{Blanco:1997a} and applied through the standard row-vector convention procedure:
\begin{equation}
\label{eq:rsh-csh}
\hat{H}^\text{(eff)}(\text{CSH}) = U^\dagger \cdot \hat{H}^\text{(eff)}(\text{RSH}) \cdot U 
\end{equation}
\noindent where $U^\dagger$ is the complex-conjugate transpose of $U$ and $\hat{H}^\text{(eff)}(\text{CSH})$ is the crystal field (effective) Hamiltonian in the CSH basis. The latter is brought in correspondence with the model matrix of $\vcf$ and each individual CFP is extracted through the ITO decomposition:
\begin{equation}
\label{eq:ito}
B_k^q(\text{Stevens}) = \frac{\mathrm{Tr}\big(\mathbf{H}^\mathrm{(eff)} \cdot \mathbf{O}_k^q \big)}{\mathrm{Tr}\big(\mathbf{O}_k^q \cdot \mathbf{O}_k^q \big)}
\end{equation}
\begin{table}[t!]
\centering
\caption{$\hat{H}^\text{CSH}$ matrix elements expressed in the $\ket{l=3,m_l}$ model basis, derived from a CASSCF calculation for the Ce$^{3+} $4f$^1$ configuration in a $D_\text{4h}$ field of point charges.$^a$}
\label{tab:heff-sr}
\begin{tabular}{rccccccc}
\toprule
$\hat{H}(\text{CSH})$ & 
$\ket{3, -3}$ & $\ket{3, -2}$ & $\ket{3, -1}$ & $\ket{3, 0}$ & $\ket{3, 1}$ & $\ket{3, 2}$ & $\ket{3, 3}$ \\
\hline
$\bra{3, -3}$ & 1336.4 & 0 & 0 & 0 & 131.5 & 0 & 0 \\
$\bra{3, -2}$ & 0 & $-$96.3 & 0 & 0 & 0 & 138.4 & 0 \\
$\bra{3, -1}$ & 0 & 0 & $-$753.3 & 0 & 0 & 0 & 131.5 \\
$\bra{3, 0}$  & 0 & 0 & 0 & $-$973.6 & 0 & 0 & 0 \\
$\bra{3, 1}$  & 131.5 & 0 & 0 & 0 & $-$753.3 & 0 & 0 \\
$\bra{3, 2}$  & 0 & 138.4 & 0 & 0 & 0 & $-$96.3 & 0 \\
$\bra{3, 3}$  & 0 & 0 & 131.5 & 0 & 0 & 0 & 1336.4 \\
\toprule
\end{tabular}
\flushleft{$^a$All values are expressed in $\icm$ units.}
\end{table}
\noindent where the boldface symbols, $\mathbf{H}^\mathrm{(eff)}$ and $\mathbf{O}_k^q$, denote the matrix representations of the corresponding operators $\hat{H}^\mathrm{(eff)}(\text{CSH})$ and $\hat{O}_k^q$ expressed in the $\ket{l=3,m_l}$ model basis set. Note that the $\mathbf{O}_k^q$ matrices already include the $a_k^l$ pre-factors. If one aims for CF parameters in the  Wybourne convention,\cite{Wybourne:1965} $B_k^q(\text{Wybourne})$, the following relation may be used for any given $X$ basis:\cite{Duros:2025a}
\begin{equation}
    \label{eq:conventions}
    B_k^q(\text{Wybourne}, X) = \frac{B_k^q (\text{Stevens}, X)}{\lambda_{kq}} = \frac{A_k^q \langle r^k \rangle (X)}{\lambda_{kq}}
\end{equation}
\noindent where $\lambda_{kq}$ are coefficients tabulated in Ref.\cite{Duros:2025a}
\begin{table}[t!]
\caption{\emph{Ab initio} CFPs, $B_k^q(\text{Stevens})$, extracted in the spin-free $\ket{l=3,m_l}$ basis describing the f$^1$ configuration, for the Ce$^{3+}$ ion in point-charge fields of various symmetries.$^{a,b}$}
\label{tab:cfps-sr}
\begin{tabular}{lccccccccc}
\toprule
CFP 
& $B_2^0$ & $B_2^2$ & $B_4^0$ & $B_4^2$ & $B_4^4$ & $B_6^0$ & $B_6^2$ & $B_6^4$ & $B_6^6$ \\
\hline
$D_\text{4h}$ & $-$1944.5 & 0.0 & 54.1 & 0 & 641.6 & $-$4.1 & 0.0 & $-$45.9 & 0 \\
  & $-$1944.7 & 0.0 & 54.1 & 0 & 638.8 & $-$4.0 & 0.0 & $-$44.0 & 0 \\
\\
$O_\text{h}$ & 0.2 & 0 & 128.5 & 0 & 642.8 & 2.0 & 0 & $-$43.2 & 0 \\
  & 0.0 & 0 & 128.5 & 0 & 642.7 & 2.1 & 0 & $-$43.1 & 0 \\
\\
$T_\text{d}$  & 0.1 & 0 & $-$60.7 & 0 & $-$303.3 & 0.2 & 0 & $-$3.8 & 0 \\
  & 0.0 & 0 & $-$60.7 & 0 & $-$303.4 & 0.2 & 0 & $-$3.9 & 0 \\
\\
$C_\text{2v}$ & $-$133.7 & 1799.3 & $-$8.5 & 35.0 & 122.1 & 0.5 & $-$5.1 & 4.4 & 8.6 \\
  & $-$135.8 & 1798.6 & $-$9.2 & 35.1 & 126.3 & 0.5 & $-$5.1 & 5.0 & 9.0 \\
\toprule
\end{tabular}
\flushleft{$^a$All values are expressed in $\icm$; $^b$For each point group, the first line reports values obtained with the present extraction procedure; the second line reports data calculated using the \texttt{SINGLE\_ANISO} code.}
\end{table}

For the particular example of a 4f$^1$ configuration in a $D_\text{4h}$ CF, the state-averaged CASSCF calculation produced an orbital non-degenerate $A_{2u}$ ground state (occupied f$_\sigma$ orbital with pseudo-axial symmetry designation), followed by two $E_u$ orbital doublets at 212.0 (f$_\pi$) and 2318.2~$\icm$ (f$_\phi$), and two orbital singlets, $B_{2u}$ at 738.9~$\icm$ (f$_{\delta_-}$) and $B_{1u}$ at 1015.7~$\icm$ (f$_{\delta_+}$). The \emph{ab initio} CF matrix in the CSH basis, shown in Table~\ref{tab:heff-sr}, is real, Hermitian, and fully consistent with the structure of the model CF matrix in Table~\ref{tab:vcf-mod}, as well as with the corresponding matrix calculated using the SINGLE\_ANISO and AILFT codes implemented in ORCA. Note that the AILFT matrix elements reported in the standard output are expressed in the RSH basis and must be transformed to the CSH basis before comparison with Table~\ref{tab:heff-sr}. The CFPs extracted via the ITO decomposition are listed in Table~\ref{tab:cfps-sr}, together with parameter values obtained for crystal-field symmetries other than $D_{\mathrm{4h}}$. Data for $D_{\infty\text{h}}$ and $C_{\text{3v}}$ are given in Table~\ref{tab:cfps-sr-si}. Notably, in all symmetry cases, the new procedure yields parameter values that are similar to those calculated with the \texttt{SINGLE\_ANISO} and AILFT codes. In the following, we present our extraction procedure in the presence of SOC, demonstrating that the CFP values remain unchanged, which is a prerequisite for a full and consistent extraction of the CFP values at this level (based on CASSCF calculations with a ``minimal'' active space of 1 electron in 7 orbitals and SOCI).

\section{\label{sec:cfps-soc}Extraction of CFPs in presence of SOC\protect\\ }

When the SOC is considered, the f$^1$ configuration of a free ion is commonly described by the coupled total angular momentum basis set, $\jmj$. The coupling between the spin ($s=1/2$) and orbital angular momentum ($l=3$) generates ground $j = 5/2$ and excited $j = 7/2$ manifolds, separated by approximately 2253~$\icm$ for a free Ce$^{3+}$ ion according to the NIST database. The splitting is much larger for a valence-isoelectronic 5f$^1$ free ion, \textit{e.g.} $\sim$7608 $\icm$ for U$^{5+}$.\cite{Kaufman:1976a, Gourier:1998a} Despite the large gap, the two $j$ manifolds may merge and even intercalate between them (overlap and intermix) or with other low-energy charge transfer states in crystalline and molecular environments. This should be interpreted as a consequence of the strong influence of the coordination sphere, especially in systems with low symmetry or significant covalent bonding, as it occur often with cerium and the early actinides. In such cases, both $j$ multiplets must be considered in CF models in order to achieve rigorous and unbiased extraction of CFPs. 

In addition to the CF potential, $\hat{V}^\text{CF}$, the total Hamiltonian modeling the energy spectrum of an f$^1$ system now also includes the spin--orbit potential, $\hat{V}^\text{SO}$. The latter, is straightforwardly derived in the complete $\jmj$ basis using the standard spin-orbit operator in the spherical approximation:

\begin{equation}
    \hat{V}^\text{SO} = \zeta_\text{4f}[\hat{L}_z\hat{S_z} + \tfrac{1}{2}(\hat{L}_+\hat{S}_- + \hat{S}_- \hat{L}_+)] = \tfrac{1}{2}\zeta_{4f} (
\hat{J}^2 - \hat{L}^2 - \hat{S}^2 )
    \label{eq:vso-operator}
\end{equation}
\noindent Here, $\zeta_\text{4f}$ denotes the effective spin–orbit splitting constant of the 4f shell. The resulting SOC matrix is diagonal in the complete $\ket{j,m_j}$ basis, with diagonal entries $-2\zeta_\text{4f}$ and $\tfrac{3}{2}\zeta_\text{4f}$ in the $j=\tfrac{5}{2}$ and $j=\tfrac{7}{2}$ blocks, respectively. Accordingly, our CF model introduces $\zeta_\text{4f}$ as an explicit parameter, which will be extracted from the \emph{ab initio} data through the ITO procedure, after we present, in the following, the construction of the model matrix of $\hat{V}^\text{CF}$ and the numerical (effective) Hamiltonian, $\hat{H}^\text{eff}$, both expressed in the complete spin–orbit basis. We note that an explicit treatment of $\zeta_\text{4f}$ is also adopted in AILFT, where its value is instead determined by fitting the \emph{ab initio} SOC matrix.\cite{Atanasov:2011AILFT, Schweinfurth:2015a,Aravena:2016a,Walisinghe:2021a}

\subsection{\label{subsec:cfps-soc-model}Construction of the model CF matrix of $\vcf$\protect\\ }

The model matrix of $\vcf$ becomes 14$\times$14-dimensional for an f$^1$ configuration, and easily constructed by expanding the operator equivalents of Equation \ref{Eq:1}, this time in the $X=j$ representation and retaining $a_k^j$ reduced matrix elements. For $j=5/2$ levels, these prefactors are $-2/35$, $2/315$ and 0 for $k$ = 2, 4 and 6, respectively. The corresponding values for the $j=7/2$ levels are $-2/63$, $2/1155$, and $-4/27027$.\cite{Abragam:1970a} To exemplify the construction of model matrix, we return to the 4f$^1$ Ce$^{3+}$ test case with $D_\text{4h}$ symmetry. The $\vcf$ expression remains that of Equation \ref{Eq:2}; it includes the same CFPs as in the spin-free case, but the matrix elements are now evaluated in the $j$ basis set. The resulting CF model matrix has the form:
\begin{equation}
\label{eq:vcf-j}
\mathbf{V^{\text{CF}}} = \begin{pmatrix}
V^\text{CF}_{j=5/2} & [V_{j,j^\prime}^{\text{CF}}]^\dagger \\
V^\text{CF}_{j,j^\prime} & V^\text{CF}_{j^\prime=7/2}
\end{pmatrix}
\end{equation}
\noindent with expressions for the two blocks on the main diagonal, $V^\text{CF}_{j=5/2}$ and $V^\text{CF}_{j^\prime=7/2}$, derived in Tables \ref{tab-si:cf-j52} and \ref{tab-si:cf-j72}. A notable feature of this (approximate) matrix is its block-diagonal structure, i.e. the off-diagonal blocks contain $V^\text{CF}_{j,j^\prime}$ which is a ${0}_{8 \times 6}$ matrix. This is because the operator equivalents in Equation \ref{Eq:1} (and implicitly in Equation \ref{Eq:2}) do not generate any couplings between the two $j=5/2$ and $j^\prime=7/2$ multiplets. In other words, by construction, the model CF matrix constructed directly in the $\jmj$ basis set cannot account for $j$-mixing. Abragam and Bleaney noted this shortcoming, and suggested instead that these couplings be derived using the Wigner–Eckart theorem within the Racah formalism.\cite{AbragamBleaney} 

The solution for adhering to the Stevens operator equivalents formalism is actually straightforward, i.e., one can construct the model matrix of $\vcf$ directly in the 14-dimensional uncoupled spin–orbital basis, $\lmlsms$. Accordingly, the expression for the CF potential is given by Equation~\ref{Eq:2} (referring to our $D_\text{4h}$ test case), and the model matrix is developed in the $l$-representation, exactly as in the spin-free case. The passage of this model matrix from the uncoupled $\lmlsms$ to the coupled total angular momentum basis, $\jmj$, proceeds through the unitary transformation matrix of CG coefficients ($U^\text{CG}$) provided in Table \ref{tab-si:ucg}:
\begin{equation}
\label{eq:ucg-rot}
\vcf(\jmj) = U^\text{CG} \vcf (\lmlsms) U^{\text{CG}\dagger}
\end{equation}
\noindent The resulting CF matrix, given by Equation~\ref{eq:vcf-j}, is identical in what concerns the two blocks on the main diagonal, but now contains nonzero off-diagonal blocks with $V^\text{CF}_{j,j^\prime}$ expressions relevant for the $j$-mixing given in Table~\ref{tab:cf-jmixing}). The CFPs should now be extracted by correspondence with a numerical (effective) CF Hamiltonian matrix, $\hat{H}^\text{(eff)}$, which is derived from \emph{ab initio} SOCI calculations in the following subsection. Note that this is analogous to describing the spin mixing in binuclear transition metal complexes in the weak-exchange limit \cite{Maurice:2010MS, Maurice:2010GS, Sergentu:2024}.

\begin{table}[t!]
\centering
\caption{Coupling elements between $j=5/2$ and $j^\prime=7/2$ multiplets in the model crystal field matrix for an f$^1$ configuration in $D_\text{4h}$ symmetry.$^{a,b}$}
\label{tab:cf-jmixing}
\begin{tabular}{ccccccc}
\toprule
$\hat{V}^\text{CF}$ & $\ket{\frac{5}{2},-\frac{5}{2}}$ & $\ket{\frac{5}{2},-\frac{3}{2}}$ & $\ket{\frac{5}{2},-\frac{1}{2}}$ & $\ket{\frac{5}{2},+\frac{1}{2}}$ & $\ket{\frac{5}{2},+\frac{3}{2}}$ & $\ket{\frac{5}{2},+\frac{5}{2}}$ \\
\hline
$\bra{\frac{7}{2},-\frac{7}{2}}$ & $0$ & $0$ & $0$ & $\thead{\frac{16 \sqrt{105} \mathbf{A}}{45045}}$ & $0$ & $0$ \\
$\bra{\frac{7}{2},-\frac{5}{2}}$ & $\thead{\frac{2 \sqrt{6} \mathbf{B}}{3003}}$ & $0$ & $0$ & $0$ & $\thead{\frac{16 \sqrt{30} \mathbf{C}}{45045}}$ & $0$ \\
$\bra{\frac{7}{2},-\frac{3}{2}}$ & $0$ & $\thead{\frac{2 \sqrt{10} \mathbf{D}}{15015}}$ & $0$ & $0$ & $0$ & $\thead{\frac{16 \sqrt{2} \mathbf{E}}{3003}}$ \\
$\bra{\frac{7}{2},-\frac{1}{2}}$ & $0$ & $0$ & $\thead{\frac{4 \sqrt{3} \mathbf{F}}{15015}}$ & $0$ & $0$ & $0$ \\
$\bra{\frac{7}{2},+\frac{1}{2}}$ & $0$ & $0$ & $0$ & $-\thead{\frac{4 \sqrt{3} \mathbf{F}}{15015}}$ & $0$ & $0$ \\
$\bra{\frac{7}{2},+\frac{3}{2}}$ & $\thead{-\frac{16 \sqrt{2} \mathbf{E}}{3003}}$ & $0$ & $0$ & $0$ & $\thead{-\frac{2 \sqrt{10} \mathbf{D}}{15015}}$ & $0$ \\
$\bra{\frac{7}{2},+\frac{5}{2}}$ & $0$ & $\thead{-\frac{16 \sqrt{30} \mathbf{C}}{45045}}$ & $0$ & $0$ & $0$ & $\thead{-\frac{2 \sqrt{6} \mathbf{B}}{3003}}$ \\
$\bra{\frac{7}{2},+\frac{7}{2}}$ & $0$ & $0$ & $\thead{-\frac{16 \sqrt{105} \mathbf{A}}{45045}}$ & $0$ & $0$ & $0$ \\
\toprule
\end{tabular}
\flushleft{$^a$ This is $V_{j,j^\prime}^\text{CF}$ in Equation \ref{eq:vcf-j}. $^b$ Expressions for the placeholders \textbf{A, B,} etc., which are linear combinations of the $\bkq$ CFPs, are given in Table \ref{tab-si:cf-jmixing-placeholders}.}
\end{table}

\subsection{\label{subsec:cfps-soc-numerical}Derivation of the CF matrix from \emph{ab initio} calculations\protect\\ }

\begin{table*}[t!]
    \small
    \centering
    \caption{\emph{Ab initio} CF coupling elements ($\icm$ units) between the $j=5/2$ and $j^\prime=7/2$ multiplets, derived from a SOCI calculation on a Ce$^{3+}$ ion in a field of point-charges with $D_\text{4h}$ symmetry.$^{a}$}
    \label{tab:hcf-jmixing}
    \renewcommand{\arraystretch}{1.3}
    \setlength{\tabcolsep}{3pt}
  \begin{tabular}{ccccccc}
    \toprule
  $\hat{H}\mathrm{(CSH)}$   & $\ket{\frac{5}{2},-\frac{5}{2}}$ & $\ket{\frac{5}{2},-\frac{3}{2}}$ & $\ket{\frac{5}{2},-\frac{1}{2}}$ & $\ket{\frac{5}{2},+\frac{1}{2}}$ & $\ket{\frac{5}{2},+\frac{3}{2}}$ & $\ket{\frac{5}{2},+\frac{5}{2}}$ \\ \hline
    $\bra{\frac{7}{2},-\frac{7}{2}}$ & $0i$ & $0i$ & $0i$ & $98.8-0.0i$ & $0i$ & $0i$ \\
    $\bra{\frac{7}{2},-\frac{5}{2}}$ & $-500.7+0.0i$ & $0i$ & $0i$ & $0i$ & $81.2-0.0i$ & $0i$ \\
    $\bra{\frac{7}{2},-\frac{3}{2}}$ & $0i$ & $-294.9-0.0i$ & $0i$ & $0i$ & $0i$ & $74.5+0.0i$ \\
    $\bra{\frac{7}{2},-\frac{1}{2}}$ & $0i$ & $0i$ & $-107.8-0.0i$ & $0i$ & $0i$ & $0i$ \\
    $\bra{\frac{7}{2},+\frac{1}{2}}$ & $0i$ & $0i$ & $0i$ & $107.8-0.0i$ & $0i$ & $0i$ \\
    $\bra{\frac{7}{2},+\frac{3}{2}}$ & $-74.5+0.0i$ & $0i$ & $0i$ & $0i$ & $294.9-0.0i$ & $0.0i$ \\
    $\bra{\frac{7}{2},+\frac{5}{2}}$ & $0i$ & $-81.2+0.0i$ & $0i$ & $0i$ & $0i$ & $500.7+0.0i$ \\
    $\bra{\frac{7}{2},+\frac{7}{2}}$ & $0i$ & $0i$ & $-98.8+0.0i$ & $0i$ & $0i$ & $0i$ \\
    \toprule
  \end{tabular}
\flushleft{$^a$For comparison, the energy separation between the $j=\tfrac{5}{2}$ and $j^\prime =\tfrac{7}{2}$ manifolds is $\tfrac{7}{2}\zeta_\text{4f}=2379$ cm$^{-1}$.}
\end{table*}

The SOCI calculation generates a set of spin-orbit-coupled wavefunctions, Kramers doublets for odd-electron configurations such as f$^1$, that are linear combinations of the spin components of the CASSCF states. The latter, are themselves expressed as linear combinations of Slater determinants in the localized f-orbitals, as described in the previous section. This time, the expansion coefficients in the spin--orbit wavefunctions are complex numbers. 

Generally, if the spin-free CASSCF states were calculated with an f$^1$ active space, then the SOCI calculation yields the 14 spin-orbit states, $\psi_k$, and their corresponding energies, $E_k$. These can be easily mapped onto the $j = 5/2$ and $j^\prime = 7/2$ multiplets. For instance, for the Ce$^{3+}$ ion in a $D_\text{4h}$ CF (our test case), the SOCI calculation based on seven spin-doublets produced seven Kramers pairs. Three of these span the energy range up to 1824.8 $\icm$, corresponding to the CF-split $j = 5/2$ manifold, while the remaining four pairs, spanning 2348.9--4580.2 $\icm$, correspond to the $j^\prime = 7/2$ multiplet. 

Similar to the spin-free case, we proceed by constructing the \( 14 \times 14 \) eigenvector matrix, where each row corresponds to a SOCI wavefunction, $\psi_k$ (or otherwise $\psi_k^L$), expressed in the uncoupled spin-orbital basis, $\lmlsms$. The wavefunction expressions in this basis are readily obtained by multiplying the $\psi_k$ (or $\psi_k^L$) expansion coefficients with the Slater determinant coefficients of the corresponding CASSCF states in the expansion. Finally, the SOCI eigenvector matrix is fed into the des Cloizeaux algorithm of Equation \ref{eq:cloizeaux}, where, this time, $E_k$ are the SOCI energies, and $\phi_i$, $\phi_j$ are model space $\ket{l=3,m_l,s=\frac{1}{2},m_s}$ spin-orbital functions. The procedure yields the numerical CF Hamiltonian matrix expressed in the uncoupled $\lmlsms$ basis set, where the $\lml$ orbital components are yet RSHs, $\mathbf{H}(\text{RSH})$, with the following structure:
\begin{widetext}
\begin{equation}
\label{eq:hso-rsh}
\mathbf{H}(\text{RSH}) =
\begin{pmatrix}
\bra{-\tfrac{1}{2}=m_s,m_l}\hat{H}\ket{m_l,m_s=-\tfrac{1}{2}} &
\bra{-\tfrac{1}{2}=m_s,m_l}\hat{H}\ket{m_l,m_s=+\tfrac{1}{2}} \\
\bra{+\tfrac{1}{2}=m_s,m_l}\hat{H}\ket{m_l,m_s=-\tfrac{1}{2}} &
\bra{+\tfrac{1}{2}=m_s,m_l}\hat{H}\ket{m_l,m_s=+\tfrac{1}{2}}
\end{pmatrix}
\end{equation}
\end{widetext}
\noindent where $m_l \in \{-3,\dots,+3\}$. The basis-transformation to CSHs is easily achieved through Equation \ref{eq:rsh-csh}, using this time a 14$\times$14 block antidiagonal matrix containing two 7$\times$7 $U$ matrices from Table~\ref{tab-si:rsh-csh} on the antidiagonal (i.e. using the $U$ RSH$\rightarrow$CSH transformation matrix of the spin-free case). The procedure yields the numerical CF Hamiltonian matrix in the $\lmlsms$ basis, but with a CSH representation for the orbital parts, \textbf{H}(CSH). This matrix is finally translated into the total angular momentum $\jmj$ basis set by using Equation \ref{eq:ucg-rot} and the matrix $U^\text{CG}$ of Table \ref{tab-si:ucg}.

Concerning our Ce$^{3+}$ $D_\text{4h}$ test case, the numerical \textbf{H}(CSH) derived from the SOCI calculation is presented across Tables \ref{tab-si:hcf-j52} and \ref{tab-si:hcf-j72}, showing the matrix elements within the $j=5/2$ and $j^\prime=7/2$ manifolds respectively, both forming the two main diagonal blocks in the complete 14$\times$14 CF matrix, and in Table \ref{tab:hcf-jmixing}, showing a counter-diagonal block with $\bra{\frac{7}{2}=j^\prime}\hat{H}\text{(CSH)}\ket{j=\frac{5}{2}}$ coupling elements (see Tables \ref{tab-si:hcf-jmixing-oh}, \ref{tab-si:hcf-jmixing-td} and \ref{tab-si:hcf-jmixing-c2v} for off-diagonal couplings obtained for the other symmetry cases). The correspondence with the model CF matrix, $\mathbf{V}^\text{CF}$ of Equation \ref{eq:vcf-j}, with expressions given in Tables \ref{tab-si:cf-j52} and \ref{tab-si:cf-j72} for matrix elements within the $j=5/2$ and $j^\prime=7/2$ manifolds, and in Table \ref{tab:cf-jmixing} for the $j-j^\prime$ couplings, is excellent. It is now clear that the \emph{ab initio} CF matrix inherits these off-diagonal couplings by construction, making it impossible to map it directly onto a model CF matrix formulated in the pure $\jmj$ basis. 

\begin{table}[t!]
\caption{\emph{Ab initio} CFPs (in $\icm$), $B_k^q(\text{Stevens})$, extracted in the complete spin-orbit $\ket{j\in\{\frac{5}{2},\frac{7}{2}\},m_j}$ basis describing the f$^1$ configuration, for the Ce$^{3+}$ ion in point-charge fields of various symmetries.$^{a,b}$}
\label{tab:cfps-so}
\setlength{\tabcolsep}{1pt}
\begin{tabular}{lcccccccccc}
\toprule
CFP 
& $B_2^0$ & $B_2^2$ & $B_4^0$ & $B_4^2$ & $B_4^4$ & $B_6^0$ & $B_6^2$ & $B_6^4$ & $\zeta_\text{4f}$ \\
\hline
$D_\text{4h}$ & $-$1944.5 & 0.0 & 54.1 & 0 & 641.6 & $-$4.1 & 0.0 & $-$45.9 & 679.60 \\
  & $-$1766.4 & 0.0 & 40.4 & 0 & 760.6 & n/a & n/a & n/a & n/a \\  
\\
$O_\text{h}$ & 0.2 & 0 & 128.5 & 0 & 642.8 & 2.0 & 0 & $-$43.2 & 678.48 \\
  & 0.4 & 0 & 127.0 & 0 & 637.7 & n/a & n/a & n/a & n/a  \\
\\
$T_\text{d}$  & 0.1 & 0 & $-$60.7 & 0 & $-$303.3 & 0.2 & 0 & $-$3.8 & 678.34 \\
  & 0 & 0 & $-$59.3 & 0 & $-$296.7 & n/a & n/a & n/a & n/a  \\
\\
$C_\text{2v}$ $^c$ & $-$133.7 & 1799.3 & $-$8.5 & 35.0 & 122.1 & 0.5 & $-$5.1 & 4.4 & 678.25 \\
  & $-$144.8 & 1777.5 & $-$4.8 & 28.9 & 103.0 & n/a & n/a & n/a & n/a  \\
\toprule
\end{tabular}
\flushleft{$^a$For each point group, the first line reports values obtained with the present extraction procedure; the second line reports data calculated using the \texttt{SINGLE\_ANISO} code. \\ $^b$AILFT and our extraction procedure provide similar CFPs and SOC constant, $\zeta_\text{4f}$. \\ $^c$Includes also $B_6^6=8.6$ cm$^{-1}$. }
\end{table}

The CFPs and the 4f spin-orbit parameter, $\zeta_\text{4f}$, were extracted using the ITO procedure and are listed in Table~\ref{tab:cfps-so} alongside values calculated with the \texttt{SINGLE\_ANISO} code, which also derives the CFPs fully \emph{ab initio}, but based on a model restricted to the ground $j = 5/2$ manifold. Specifically, we note that the values extracted in the complete spin-orbit basis describing the $f^1$ configuration, $\ket{j \in \{\tfrac{5}{2}, \tfrac{7}{2}\}, m_j}$, are \emph{identical} to those listed in Table~\ref{tab:cfps-sr}, obtained in the spin-free $\ket{l = 3, m_l}$ basis, and also identical with values (CFPs and $\zeta_\text{4f}$) derived through AILFT. That is, once the model is complete, the CFPs are identical regardless of the choice of basis representation. Furthermore, restricting the CF model to only the lower-energy $j = 5/2$ manifold (i) excludes the description of 6$^{\text{th}}$-rank CF terms, and (ii) introduces a bias in the extracted values that is likely to increase with the strength of the CF. In particular, such a truncated model cannot describe any CF-induced mixing between $m_j$ components of $j$ and $j'$ manifolds, with repercussions on the proper description of the GS anisotropy in cases where such admixture is strong. Briefly, in the following subsection, we show how the $j-j'$ CF admixture naturally arises within a complete CF model for the moderate case of the cerocene anion.

\section{\label{sec:cfps-cerocene}CF description of the cerocene anion\protect\\ }

\begin{table*}[t!]
\small
\caption{\emph{Ab initio} CFPs (in $\icm$), $B_k^q(\text{Stevens})$, effective SOC constant, $\zeta_{\text{4f}}$, and composition of the GS, calculated for the cerocene anion.$^a$}
\label{tab:cfps-cero}
\renewcommand{\arraystretch}{1.5}
\begin{tabular}{lccccrrcccc}
\toprule
Manifold 
& $B_2^0$ & $B_4^0$ & $B_6^0$ & $\zeta_\text{4f}$ & GS composition & $g_\parallel^c$ & $g_\perp^c$  \\
\hline
$\lml$ & $-$364.2 & $-$697.3 & 54.3 & n/a & 100\% $\ket{l=3,m_l=0}$ & n/a & n/a \\

$\ket{j=\frac{5}{2},m_j}$  & $-$374.2 & $-$541.8 & n/a & n/a & 100\% $\ket{j=\frac{5}{2},m_j=\pm1/2}$ & 0.800 & 2.400 \\

$\ket{j\in\{\frac{5}{2},\frac{7}{2}\},m_j}$ & $-$364.2 & $-$697.3 & 54.3 & 658.75  & 98.6\% $\ket{j=\frac{5}{2},m_j=\pm \frac{1}{2}}$ & 1.019 & 2.202 \\
                                            &          &          &      & &    + 1.4\% $\ket{j=\frac{7}{2},m_j=\pm \frac{1}{2}}$  \\ 

\\
$\lml ^b$ & $-$364.2 & $-$697.3 & 54.3 & & 100\% $\ket{l=3,m_l=0}$ & n/a & n/a  \\
$\ket{j=\frac{5}{2},m_j} ^b$  & $-$374.2 & $-$541.8 & n/a & & 100\% $\ket{j=\frac{5}{2},m_j=\pm1/2}$ & 1.087 & 2.328 \\
\toprule
\end{tabular}
\flushleft{$^a$AILFT and our extraction procedures provide similar CFPs and SOC constant, $\zeta_\text{4f}$. \\$^b$CF data calculated using the \texttt{SINGLE\_ANISO} code. \\$^c$The $g$ values are calculated using the expectation values for the spin and orbital angular momentum from the SOCI wavefunction. Experimental values: $g_\parallel=1.123$ and $g_\perp = 2.272$.\cite{Walter:2009a}}
\end{table*}

The cerocene anion, \ce{Ce(C8H8)2^-}, conveniently prepared as [\ce{Li(thf)4}][\ce{Ce(C8H8)2}],\cite{Kilimann:1994a,Walter:2009a} provides with a nice playground for testing CF models and has been already the subject to quite advanced CF treatments.\cite{Gendron:2015a, Cespedes:2017a, Gendron:2018a} The complex  shows high-$D_\text{8h}$ symmetry and a Ce$^{3+}$-4f$^1$ metal center that bonds only weakly with the annulene ligands through the ($m_l=\pm2$) 4f$_\delta$ orbitals, where the subscript $\delta$ gives the nodal-pattern around the main symmetry axis. The GS of the complex derives 98-99\% from the $m_j=\pm\frac{1}{2}$ levels of the $^2F_{5/2}$ free-ion term, which is due to population of the 4f$_\sigma$ and 4f$_\pi$ orbitals, and 1-2\% from the $m_j=\pm\frac{1}{2}$ components of the upper $^2F_{7/2}$ manifold. The admixture is sufficient to influence the GS anisotropy, described by a $g$-tensor with $g_\parallel=1.123$ and $g_\perp=2.272$.\cite{Walter:2009a}

The present CASSCF calculation for the 4f$^1$ configuration produced spin-free energies at 0 ($A_u$, f$_\sigma$), 540 ($E_{1u}$, f$_\pi$), 732 ($E_{3u}$, f$_\phi$) and 2250 ($E_{2u}$, f$_\delta$) $\icm$ (see Table \ref{tab-si:energies-casscf}). With SOC included, the $j=5/2$ Kramers doublets occur at 0, 591, and 1160 $\icm$, and the $j=7/2$ Kramers doublets occur at 2281, 2752, 3371 and 4146 $\icm$ (see Table \ref{tab-si:energies-casscf-so}). The energies for the GS manifold corroborate with literature data within differences in computational details, e.g.\ 0, 565, 1165 $\icm$ in Ref.\cite{Cespedes:2017a}, 0, 561, 1040 $\icm$ in Ref.\cite{Gendron:2015a}, and 0, 610, 1146 $\icm$ in Ref.\cite{Gendron:2018a}. The separation to the first excited doublets is sufficiently large in determining a quite linear behavior of the $\chi T$ magnetic susceptibility curve above 10 K, representative of second-order Zeeman interaction. Considering the pseudo-axial symmetry of this complex, the CF potential can be expressed using only the axial ($q=0$) terms:
\begin{equation}
    \vcf(X) = a_2^X [B_2^0\hat{O}_2^0(X)]+a_4^X[(B_4^0\hat{O}_4^0(X)]+a_6^X[(B_6^0\hat{O}_6^0(X)]
    \label{Eq:vpot-cero}
\end{equation}

The extracted CFPs in the spin-free $X=l$ basis and in the complete spin-orbit $X=j\in{\{\tfrac{5}{2},\tfrac{7}{2}\}}$ basis, using the present approach, are provided in Table \ref{tab:cfps-cero}, together with counterparts extracted with the \texttt{SINGLE\_ANISO} code. In particular, it is noted that the extracted values are identical when obtained within the spin-free $X = l$ representation, regardless of the used computational approach. These values also remain unchanged when evaluated in the complete $j \in \{\frac{5}{2}, \frac{7}{2}\}$ manifold. However, deviations appear when the extraction is performed using a model truncated to the $j = \frac{5}{2}$ subspace. Therefore, these differing values should not be interpreted as effects arising from the chosen basis representation or from SOC; rather, they are biased results caused by truncation of the model basis set. 

Finally, one notes in Table \ref{tab:cfps-cero} that our CF model developed in the complete $j$-basis quantitatively describes the GS admixture of excited $\ket{\frac{7}{2},\pm \frac{1}{2}}$ into the $\ket{\frac{5}{2},\pm \frac{1}{2}}$ GS, and likewise the effective $g$-tensor, supporting the experimental $g$ values and interpretation.\cite{Walter:2009a} One notes that the $g$ factors calculated with the CF model accounting only for the $j=\tfrac{5}{2}$ ground manifold ($g_\parallel=0.80$ and $g_\perp = 2.40$) are rather close to values expected for a pure axial $\ket{\tfrac{5}{2}, \pm \tfrac{1}{2}}$ level ($g_\parallel = \tfrac{6}{7}=0.86$ and $g_\perp = \tfrac{18}{7}=2.56$).\cite{Walter:2009a,Gendron:2018a} It is thus clear that even a slight $j-j^\prime$ mixing may have a large impact on properties, and such mixing should be included in a CF model.

Table \ref{tab:hcf-jmixing-cerocene} reports the numerical value of the $j-j^\prime$ CF mixing elements. One notes that, prior to matrix diagonalization, the largest off-diagonal CF matrix element is $V=770.5$ cm$^{-1}$. In comparison, the spin--orbit gap between the two manifolds is 
$\Delta_{\mathrm{SO}}=\tfrac{7}{2}\zeta_{4f}=2306$ cm$^{-1}$. 
The ratio between the largest CF mixing term and $\Delta_{\mathrm{SO}}$ is therefore 0.33. 
Within a simplified two-level assumption, this would correspond to a mixing amplitude of order 
$V/\Delta_{\mathrm{SO}}\approx 0.33$, i.e.\ a probability admixture of roughly 
$(V/\Delta_{\mathrm{SO}})^2\approx 10\%$ before full matrix diagonalization. 
This indicates that the ligand--4f interaction in the cerocene anion is sufficiently strong to partially compete with SOC. Although the $j=\tfrac{7}{2}$ contribution to the ground Kramers doublet is reduced to only 1.4\% after full matrix diagonalization, the sizable off-diagonal CF matrix elements (reaching up to $\sim\tfrac{1}{3}\Delta_{\mathrm{SO}}$) place the system in the intermediate-coupling regime. This intermediate character accounts for the large deviation of the $g$ values from those of a pure $\ket{\tfrac{5}{2},\pm\tfrac{1}{2}}$ level in a an axial CF.

It is shown, therefore, that the resolution of a CF model built in the complete $j$-basis of the $l=3$ manifold for the cerocene anion, leads to magnetic properties that naturally inherit the $j=7/2$ mixing into the $j=5/2$ ground state. This work complements the CF endeavor of Ref.\cite{Gendron:2018a}, where expressions containing the $j$-mixing are firstly derived for magnetic properties, e.g. for $g_\parallel$ and $g_\perp$, and then equated with numerical data inferred from the \emph{ab initio} wavefunctions and energies.

\begin{table*}[t!]
    \small
    \centering
    \caption{\emph{Ab initio} CF coupling elements ($\icm$ units) between the $j=5/2$ and $j^\prime=7/2$ multiplets, derived from a SOCI calculation for the cerocene anion.$^{a}$}
    \label{tab:hcf-jmixing-cerocene}
    \renewcommand{\arraystretch}{1.3}
    \setlength{\tabcolsep}{3pt}
  \begin{tabular}{ccccccc}
    \toprule
  $\hat{H}\mathrm{(CSH)}$   & $\ket{\frac{5}{2},-\frac{5}{2}}$ & $\ket{\frac{5}{2},-\frac{3}{2}}$ & $\ket{\frac{5}{2},-\frac{1}{2}}$ & $\ket{\frac{5}{2},+\frac{1}{2}}$ & $\ket{\frac{5}{2},+\frac{3}{2}}$ & $\ket{\frac{5}{2},+\frac{5}{2}}$ \\ \hline
$\bra{\frac{7}{2},-\frac{7}{2}}$ & $0i$ & $0i$ & $0i$ & $0i$ & $0i$ & $0i$ \\
$\bra{\frac{7}{2},-\frac{5}{2}}$ & $530.5 + 0.0i$ & $0i$ & $0i$ & $0i$ & $0i$ & $0i$ \\
$\bra{\frac{7}{2},-\frac{3}{2}}$ & $0i$ & $-770.5 - 0.0i$ & $0i$ & $0i$ & $0i$ & $0i$ \\
$\bra{\frac{7}{2},-\frac{1}{2}}$ & $0i$ & $0i$ & $-264.6 + 0.0i$ & $0i$ & $0i$ & $0i$ \\
$\bra{\frac{7}{2},+\frac{1}{2}}$ & $0i$ & $0i$ & $0i$ & $264.6 + 0.0i$ & $0i$ & $0i$ \\
$\bra{\frac{7}{2},+\frac{3}{2}}$ & $0i$ & $0i$ & $0i$ & $0i$ & $770.5 + 0.0i$ & $0i$ \\
$\bra{\frac{7}{2},+\frac{5}{2}}$ & $0i$ & $0i$ & $0i$ & $0i$ & $0i$ & $-530.5 - 0.0i$ \\
$\bra{\frac{7}{2},+\frac{7}{2}}$ & $0i$ & $0i$ & $0i$ & $0i$ & $0i$ & $0i$ \\
    \toprule
  \end{tabular}
\flushleft{$^a$For comparison, the energy separation between the $j=\tfrac{5}{2}$ and $j^\prime =\tfrac{7}{2}$ manifolds is $\tfrac{7}{2}\zeta_\text{4f}=2306$ cm$^{-1}$.}
\end{table*}
 
\section{\label{sec:conclusion}Conclusion\protect\\ }

This study presents a step-by-step strategy for describing the f$^1$ configuration of lanthanide and actinide ions, fully \emph{ab initio}, within the framework of crystal field (CF) theory, the widely used Stevens operator-equivalent formalism, and quasi-relativistic and multiconfiguration wavefunction theory calculations. The study addresses a potential confusion in the field--namely, whether spin-orbit coupling (SOC) influences the values of the crystal field parameters (CFPs), and whether such effects arise from the limitations of traditional approaches that rely on truncated spin-orbit manifolds, $\JMJ$, stemming from the ground orbital term ($L$). 

Showcasing with a Ce$^{3+}$ 4f$^1$ configuration in various CFs of point charges, we were able to demonstrate that the extracted CFPs are \emph{identical} when a complete model interaction space is employed, e.g. the complete set of $\ket{l=3,m_l}$ model functions when the parameters are extracted in the spin-free formalism, or the complete set of $\ket{j\in \{\frac{5}{2},\frac{7}{2}\},m_j}$, when the parameters are extracted in the spin-orbit-coupled formalism. In this latter case, truncation of the model space to the ground $j=5/2$ manifold, introduces a bias in the extracted parameters that is likely to increase with the strength of the CF. This study demonstrates for the 4f$^1$ case that parameters extracted within the parent $L$-manifold are valid for describing the splitting across all $J$-manifolds derived therefrom, provided that the effective spin--orbit constant is treated explicitly in the model.

For all the 4f$^1$ cases studied in this work, the extracted values of the CFPs and the effective spin-orbit constant, $\zeta_\text{4f}$, were similar to those obtained from \emph{ab initio} ligand field theory (AILFT). This comparison is reassuring, since AILFT extracts parameters directly at the orbital level rather than from the coupled $\ket{j,m_j}$ states, and no additional parameters need to be introduced in the CF model for the f$^1$ case. A limitation is that, since both our extraction procedure and AILFT use the localized 4f orbitals generated by AILFT, which may not be 100\% pure, the error on the SOC treatment in AILFT--ranging from roughly 0.5 to 8 cm$^{-1}$ in certain matrix elements of the effective CF Hamiltonian derived from the \emph{ab initio} data--is also present in our extraction. This type of error cannot be easily mitigated, since even a small contribution from, e.g., 5p basis functions to the 4f-localized orbitals will appear in the spin-orbit matrix-contributions that cannot be recovered without increasing the complexity of the CF model.

In summary, the present study provides with a novel CF model extraction for the $f^1$ case in the complete coupled $j$-basis, through the \emph{ab initio} (effective) Hamiltonian theory. This work can be viewed as a complement to the AILFT extraction, where the CF Hamiltonian matrix is derived through a fitting procedure. It provides a robust and systematic framework for deriving CFPs from first principles and offers a new perspective for modeling the electronic structures of lanthanide and actinide systems. The approach is expected to have broad applications in the design of molecular magnets, luminescent materials, and quantum technologies. Work is currently in progress to report CF data for $f^1$ systems beyond the simple case of the cerocene anion presented herein, as well as for the $f^2$ case, for which the \emph{NewMag} code is currently under development. Note that starting with the f$^2$ configuration, we expect deviations with the AILFT procedure, because $L$ and $l$ will not match anymore as in the f$^1$ configuration. 

Finally, unlike the lanthanide 4f orbitals, the 5f orbitals of the actinides can make a significant contribution to metal--ligand bonding. Consequently, in actinide complexes, spin--orbit--induced orbital relaxation effects may become important and the CFPs can be affected by SOC, giving rise to what is known as the SOC-induced relativistic crystal-field effect seen in transition-metal cases .~\cite{Atanasov:2005a}

\section*{Conflict of Interest}
\noindent The authors have no conflicts to disclose.

\begin{acknowledgments}
D.-C.S. and I.H. acknowledge mobility funding provided through the Romania–France bilateral research program, supported by the Romanian National Authority for Scientific Research and Innovation (UEFISCDI), project code PN-IV-P8-8.3-PM-RO-FR-2024-00-22. G.D.-R., B.L.G. and R.M. acknowledge the ''PHC Brancusi'' program (project number: 51686YA), funded by the French Ministry for Europe and Foreign Affairs, the French Ministry for Higher Education and Research and the Ministry of Research, Innovation and Digitalization (M.C.I.D.). Additional support by the ANR (Contract No. ANR-23-PETQ-0007) is also acknowledged.
\end{acknowledgments}

\section*{Data Availability Statement}

The data supporting the findings of this study are available within the article, its supplementary material, and the \texttt{NewMag} code repository, which is publicly available on GitHub at \texttt{https://github.com/clausserg/newmag.git}. The \texttt{examples} folder in this repository contains all raw output from the \emph{ab initio} calculations reported in this manuscript.

\section*{Supplementary Material}
\noindent The supplementary material reports complimentary tables indented in this article, calculated \emph{ab initio} energies, and $xyz$ Cartesian coordinates for the cerocene anion and \ce{Ce^{3+}} point-charge models.

\section*{REFERENCES}
\bibliography{references}

\appendix

\section*{Supplementary Material}

\renewcommand{\thepage}{S\arabic{page}} 
\renewcommand{\thesection}{S\arabic{section}}  
\renewcommand{\thetable}{S\arabic{table}}  
\renewcommand{\thefigure}{S\arabic{figure}}
\setcounter{page}{1}
\setcounter{figure}{0}
\setcounter{table}{0}

\begin{table*}[h!]
  \centering
  \caption{ XYZ cartesian coordinates ($\text{Å}$) for the cerocene anion.}
  \label{tab-si:cerocene_coords}
  \begin{tabular}{lrrr}
    \toprule
    Atom & \multicolumn{1}{c}{X} & \multicolumn{1}{c}{Y} & \multicolumn{1}{c}{Z} \\ \hline

    Ce & $0.000000000000$ & $0.000000000000$ & $0.000000000000$ \\
    C  & $1.834414297760$ & $0.000000000000$ & $-2.040000000000$ \\
    C  & $1.297126789452$ & $-1.297126789452$ & $-2.040000000000$ \\
    C  & $0.000000000000$ & $-1.834414297760$ & $-2.040000000000$ \\
    H  & $2.914735082122$ & $0.000000000000$ & $-1.956467000000$ \\
    H  & $2.061028941931$ & $-2.061028941931$ & $-1.956467000000$ \\
    H  & $0.000000000000$ & $-2.914735082122$ & $-1.956467000000$ \\
    C  & $-1.834414297760$ & $0.000000000000$ & $-2.040000000000$ \\
    C  & $1.834414297760$ & $0.000000000000$ & $2.040000000000$ \\
    C  & $-1.834414297760$ & $0.000000000000$ & $2.040000000000$ \\
    C  & $-1.297126789452$ & $1.297126789452$ & $-2.040000000000$ \\
    C  & $1.297126789452$ & $1.297126789452$ & $2.040000000000$ \\
    C  & $-1.297126789452$ & $-1.297126789452$ & $2.040000000000$ \\
    C  & $-1.297126789452$ & $1.297126789452$ & $2.040000000000$ \\
    C  & $1.297126789452$ & $-1.297126789452$ & $2.040000000000$ \\
    C  & $-1.297126789452$ & $-1.297126789452$ & $-2.040000000000$ \\
    C  & $1.297126789452$ & $1.297126789452$ & $-2.040000000000$ \\
    C  & $0.000000000000$ & $1.834414297760$ & $-2.040000000000$ \\
    C  & $0.000000000000$ & $1.834414297760$ & $2.040000000000$ \\
    C  & $0.000000000000$ & $-1.834414297760$ & $2.040000000000$ \\
    H  & $-2.914735082122$ & $0.000000000000$ & $-1.956467000000$ \\
    H  & $2.914735082122$ & $0.000000000000$ & $1.956467000000$ \\
    H  & $-2.914735082122$ & $0.000000000000$ & $1.956467000000$ \\
    H  & $-2.061028941931$ & $2.061028941931$ & $-1.956467000000$ \\
    H  & $2.061028941931$ & $2.061028941931$ & $1.956467000000$ \\
    H  & $-2.061028941931$ & $-2.061028941931$ & $1.956467000000$ \\
    H  & $-2.061028941931$ & $2.061028941931$ & $1.956467000000$ \\
    H  & $2.061028941931$ & $-2.061028941931$ & $1.956467000000$ \\
    H  & $-2.061028941931$ & $-2.061028941931$ & $-1.956467000000$ \\
    H  & $2.061028941931$ & $2.061028941931$ & $-1.956467000000$ \\
    H  & $0.000000000000$ & $2.914735082122$ & $-1.956467000000$ \\
    H  & $0.000000000000$ & $2.914735082122$ & $1.956467000000$ \\
    H  & $0.000000000000$ & $-2.914735082122$ & $1.956467000000$ \\ \hline
    \toprule
  \end{tabular}
\end{table*}

\begin{table*}[h!]
  \centering
  \caption{ XYZ cartesian coordinates ($\text{Å}$) for Ce$^{3+}$ point-charge models.}
  \label{tab-si:model_coords}
  \begin{tabular}{lcrrr}
    \toprule
    $D_{4h}$ & Charge & \multicolumn{1}{c}{X} & \multicolumn{1}{c}{Y} & \multicolumn{1}{c}{Z} \\ \hline
Ce &     &  0.000000000000 &  0.000000000000 &  0.000000000000 \\
Q  & -2  &  3.000000000000 &  0.000000000000 &  0.000000000000 \\
Q  & -2  & -3.000000000000 &  0.000000000000 &  0.000000000000 \\
Q  & -2  &  0.000000000000 &  3.000000000000 &  0.000000000000 \\
Q  & -2  &  0.000000000000 & -3.000000000000 &  0.000000000000 \\ \hline 
    $O_{h}$ & Charge & \multicolumn{1}{c}{X} & \multicolumn{1}{c}{Y} & \multicolumn{1}{c}{Z} \\ \hline
Ce &     &  0.000000000000 &  0.000000000000 &  0.000000000000 \\
Q  & -2  & -3.000000000000 &  0.000000000000 &  0.000000000000 \\
Q  & -2  &  3.000000000000 &  0.000000000000 &  0.000000000000 \\
Q  & -2  &  0.000000000000 &  0.000000000000 &  3.000000000000 \\
Q  & -2  &  0.000000000000 &  0.000000000000 & -3.000000000000 \\
Q  & -2  &  0.000000000000 & -3.000000000000 &  0.000000000000 \\
Q  & -2  &  0.000000000000 &  3.000000000000 &  0.000000000000 \\ \hline 
    $T_{d}$ & Charge & \multicolumn{1}{c}{X} & \multicolumn{1}{c}{Y} & \multicolumn{1}{c}{Z} \\ \hline
Ce &     &  0.000000000000 &  0.000000000000 &  0.000000000000 \\
Q  & -2  &  1.732050807569 & -1.732050807569 &  1.732050807569 \\
Q  & -2  & -1.732050807569 &  1.732050807569 &  1.732050807569 \\
Q  & -2  & -1.732050807569 & -1.732050807569 & -1.732050807569 \\
Q  & -2  &  1.732050807569 &  1.732050807569 & -1.732050807569 \\ \hline 
    $C_{2v}$ & Charge & \multicolumn{1}{c}{X} & \multicolumn{1}{c}{Y} & \multicolumn{1}{c}{Z} \\ \hline
Ce &     &  0.000000000000 &  0.000000000000 &  0.000000000000 \\
Q  & -2  & -2.372068721236 &  0.000000000000 &  1.836651840097 \\
Q  & -2  &  2.372068721236 &  0.000000000000 &  1.836651840097 \\ \hline 
    $C_{3v}$ & Charge & \multicolumn{1}{c}{X} & \multicolumn{1}{c}{Y} & \multicolumn{1}{c}{Z} \\ \hline
Ce &     &  0.000000000000 &  0.000000000000 &  0.000000000000 \\
Q  & -2  &  1.399622344511 &  2.424217012102 & -1.078994518513 \\
Q  & -2  &  1.399622344511 & -2.424217012102 & -1.078994518513 \\
Q  & -2  & -2.799244689022 &  0.000000000000 & -1.078994518513 \\
    \toprule
  \end{tabular}
\end{table*}

\begin{table*}[h!]
\centering
\caption{CASSCF relative energies (in cm$^{-1}$) for the cerocene anion and Ce$^{3+}$ point-charge models from Table \ref{tab-si:model_coords}.}
\label{tab-si:energies-casscf}
\begin{tabular}{l|llllllll}
\toprule
Ce(C$_8$H$_8$)$_2^{-}$ & \multicolumn{5}{c}{Models} \\  \hline
$D_{8h}$ & $D_{4h}$ & $O_{h}$ & $T_{d}$ & $C_{2v}$ & $C_{3v}$ \\ \hline
0 ($A_{u}$)      & 0 ($A_{2u}$)    & 0 ($A_{2u}$)     & 0 ($T_{2}$) & 0 ($B_{2}$) & 0 ($A_{1}$) \\
539.6 ($E_{1u}$) & 212.0 ($E_{u}$) & 279.4 ($T_{2u}$) & 0 ($T_{2}$) & 3.4 ($A_{1}$) & 119.1 ($E$) \\
539.6 ($E_{1u}$) & 212.0 ($E_{u}$) & 279.4 ($T_{2u}$) & 0 ($T_{2}$) & 542.6 ($B_{1}$) & 119.1 ($E$) \\
731.7 ($E_{3u}$) & 738.9 ($B_{2u}$) & 279.4 ($T_{2u}$) & 115.8 ($T_{1}$) & 632.8 ($A_{2}$) & 495.4 ($E$)  \\
731.7 ($E_{3u}$) & 1015.7 ($B_{1u}$) & 550.2 ($T_{1u}$) & 115.8 ($T_{1}$) & 834.4 ($B_{2}$) & 495.4 ($E$) \\
2250.0 ($E_{2u}$) & 2318.2 ($E_{u}$) & 550.2 ($T_{1u}$) & 115.8 ($T_{1}$) & 1259.3 ($A_{1}$) & 1263.1 ($A_2$) \\
2250.0 ($E_{2u}$) & 2318.2 ($E_{u}$) & 550.2 ($T_{1u}$) & 265.7 ($S_{2}$) & 1271.0 ($B_{1}$) & 1271.4 ($A_1$) \\
\toprule
\end{tabular}
\end{table*}

\begin{table*}[h!]
\centering
\caption{CASSCF+SOC relative energies (in cm$^{-1}$) for the cerocene anion and Ce$^{3+}$ point-charge models from Table \ref{tab-si:model_coords}.$^a$}
\label{tab-si:energies-casscf-so}
\begin{tabular}{lrrrrrr}
\toprule
Ce(C$_8$H$_8$)$_2^{-}$ & \multicolumn{5}{c}{Models} \\  \hline
$D_{8h}$ & $D_{4h}$ & $O_{h}$ & $T_{d}$ & $C_{2v}$ & $C_{3v}$ \\ \hline
0.00    & 0.00    & 0.00    & 0.00    & 0.00    & 0.00    \\
0.00    & 0.00    & 0.00    & 0.00    & 0.00    & 0.00    \\
591.30  & 519.12  & 290.78  & 0.00    & 589.82  & 308.88  \\
591.30  & 519.12  & 290.78  & 0.00    & 589.82  & 308.88  \\
1160.38 & 1824.77 & 290.78  & 135.65  & 1084.91 & 1045.81 \\
1160.38 & 1824.77 & 290.78  & 135.65  & 1084.91 & 1045.81 \\ \hline
2280.80 & 2348.85 & 2346.11 & 2332.66 & 2296.33 & 2357.12 \\
2280.80 & 2348.85 & 2346.11 & 2332.66 & 2296.33 & 2357.12 \\
2751.57 & 2681.67 & 2604.48 & 2408.97 & 2820.50 & 2544.40 \\
2751.57 & 2681.67 & 2604.48 & 2408.97 & 2820.50 & 2544.40 \\
3371.21 & 3524.81 & 2604.48 & 2408.97 & 3150.28 & 2950.65 \\
3371.21 & 3524.81 & 2604.48 & 2408.97 & 3150.28 & 2950.65 \\
4146.05 & 4580.19 & 2772.11 & 2538.39 & 3558.66 & 3577.37 \\
4146.05 & 4580.19 & 2772.11 & 2538.39 & 3558.66 & 3577.37 \\
\toprule
\end{tabular}
\\$^a$A horizontal line is drawn between the $j=5/2$ and $j=7/2$ manifolds.
\end{table*}

\begin{table*}[h!]
\centering
\caption{Matrix $U$ used to basis-transform a CF Hamiltonian between real (RSH) and complex spherical harmonics bases (CSH).$^a$}
\label{tab-si:rsh-csh}
\begin{tabular}{rccccccc}
\toprule
$U$ & $\ket{3,-3}$ & $\ket{3,-2}$ & $\ket{3,-1}$ & $\ket{3,0}$ & $\ket{3,1}$ & $\ket{3,2}$ & $\ket{3,3}$ \\
\hline
$\bra{3, \phi_s}$  & $\frac{\sqrt{2} i}{2}$ & 0 & 0 & 0 & 0 & 0 & $\frac{\sqrt{2}}{2}$ \\
$\bra{3, \delta_s}$ & 0 & $\frac{\sqrt{2} i}{2}$ & 0 & 0 & 0 & $\frac{\sqrt{2}}{2}$ & 0 \\
$\bra{3, \pi_s}$    & 0 & 0 & $\frac{\sqrt{2} i}{2}$ & 0 & $\frac{\sqrt{2}}{2}$ & 0 & 0 \\
$\bra{3, \sigma}$   & 0 & 0 & 0 & 1 & 0 & 0 & 0 \\
$\bra{3, \pi_c}$    & 0 & 0 & $\frac{\sqrt{2} i}{2}$ & 0 & $-\frac{\sqrt{2}}{2}$ & 0 & 0 \\
$\bra{3, \delta_c}$ & 0 & $-\frac{\sqrt{2} i}{2}$ & 0 & 0 & 0 & $\frac{\sqrt{2}}{2}$ & 0 \\
$\bra{3, \phi_c}$   & $\frac{\sqrt{2} i}{2}$ & 0 & 0 & 0 & 0 & 0 & $-\frac{\sqrt{2}}{2}$ \\
\toprule
\end{tabular}
\\ $^a$RSHs on rows, CSHs on columns. Subscripts s and c denote sine and cosine components.
\end{table*}

\begin{table*}[h!]
\caption{Calculated CFPs for the 4f$^1$ configuration in various crystal fields of point charges.$^{a}$}
\label{tab:cfps-sr-si}
\begin{tabular}{lcccccccc}
\toprule
CFP ($\si{\per\centi\meter}$) 
& $B_2^0$ & $B_4^0$ & $|B_4^3|$ & $B_6^0$ & $|B_6^3|$ & $B_6^6$ \\
\hline
$D_{\infty\text{h}}$& 2012.4 & 72.9 & 0 & 5.9 & 0 & 0 \\
aniso$^b$ & 2012.4 & 72.9 & 0 & 5.9 & 0 & 0 \\
\\
$C_{3\text{v}}$ & $-$1064.2 & 24.9 & 577.9 & $-$0.7 & 24.4 & 6.0 \\
aniso$^b$ & $-$1064.1 & 25.0 & 587.6 & $-$0.7 & 25.0 & 5.9 \\
\toprule
\end{tabular}
\\ $^a$All values are expressed in $\icm$; $^b$Data calculated with the \texttt{SINGLE\_ANISO} program.
\end{table*}

\begin{table*}[h!]
\centering
\caption{Model CF matrix elements expressed within the ground $\ket{j=5/2,m_j}$ manifold of an f$^1$ configuration in a $D_\text{4h}$ crystal field. This is $V^\text{CF}_{j=5/2}$ of Equation \ref{eq:vcf-j}.}
\label{tab-si:cf-j52}

\begin{tabular}{ccccccc}
\toprule
$\vcf$ & $\mathbf{- 5/2}$ & $\mathbf{- 3/2}$ & $\mathbf{- 1/2}$ & $\mathbf{+ 1/2}$ & $\mathbf{+ 3/2}$ & $\mathbf{+ 5/2}$ \\
\hline
$\mathbf{- 5/2}$ & $- \frac{4 B^{0}_{2}}{7} + \frac{8 B^{0}_{4}}{21}$ & $0$ & $0$ & $0$ & $\frac{8 \sqrt{5} B^{4}_{4}}{105}$ & $0$ \\
$\mathbf{- 3/2}$ & $0$ & $\frac{4 B^{0}_{2}}{35} - \frac{8 B^{0}_{4}}{7}$ & $0$ & $0$ & $0$ & $\frac{8 \sqrt{5} B^{4}_{4}}{105}$ \\
$\mathbf{- 1/2}$ & $0$ & $0$ & $\frac{16 B^{0}_{2}}{35} + \frac{16 B^{0}_{4}}{21}$ & $0$ & $0$ & $0$ \\
$\mathbf{+ 1/2}$ & $0$ & $0$ & $0$ & $\frac{16 B^{0}_{2}}{35} + \frac{16 B^{0}_{4}}{21}$ & $0$ & $0$ \\
$\mathbf{+ 3/2}$ & $\frac{8 \sqrt{5} B^{4}_{4}}{105}$ & $0$ & $0$ & $0$ & $\frac{4 B^{0}_{2}}{35} - \frac{8 B^{0}_{4}}{7}$ & $0$ \\
$\mathbf{+ 5/2}$ & $0$ & $\frac{8 \sqrt{5} B^{4}_{4}}{105}$ & $0$ & $0$ & $0$ & $- \frac{4 B^{0}_{2}}{7} + \frac{8 B^{0}_{4}}{21}$ \\ 
\toprule
\end{tabular}
\end{table*}

\begin{table*}[h!]
\centering
\caption{Model CF matrix elements expressed within the ground $\ket{j=7/2,m_j}$ manifold of an f$^1$ configuration in a $D_\text{4h}$ crystal field. This is $V^\text{CF}_{j^\prime=7/2}$ of Equation \ref{eq:vcf-j}.}
\label{tab-si:cf-j72}
\renewcommand{\arraystretch}{1.3}
\setlength{\tabcolsep}{3pt}

\begin{tabular}{ccccc}
\toprule
$\vcf$ & $\mathbf{- 7/2}$ & $\mathbf{- 5/2}$ & $\mathbf{- 3/2}$ & $\mathbf{- 1/2}$ \\
\hline
$\mathbf{- 7/2}$ & $- \frac{2 B^{0}_{2}}{3} + \frac{8 B^{0}_{4}}{11} - \frac{80 B^{0}_{6}}{429}$ & $0$ & $0$ & $0$ \\
$\mathbf{- 5/2}$ & $0$ & $- \frac{2 B^{0}_{2}}{21} - \frac{104 B^{0}_{4}}{77} + \frac{400 B^{0}_{6}}{429}$ & $0$ & $0$ \\
$\mathbf{- 3/2}$ & $0$ & $0$ & $\frac{2 B^{0}_{2}}{7} - \frac{24 B^{0}_{4}}{77} - \frac{240 B^{0}_{6}}{143}$ & $0$ \\
$\mathbf{- 1/2}$ & $0$ & $0$ & $0$ & $\frac{10 B^{0}_{2}}{21} + \frac{72 B^{0}_{4}}{77} + \frac{400 B^{0}_{6}}{429}$ \\
$\mathbf{+ 1/2}$ & $\frac{8 \sqrt{35} B^{4}_{4}}{385} - \frac{80 \sqrt{35} B^{4}_{6}}{3003}$ & $0$ & $0$ & $0$ \\
$\mathbf{+ 3/2}$ & $0$ & $\frac{8 \sqrt{3} B^{4}_{4}}{77} + \frac{80 \sqrt{3} B^{4}_{6}}{1287}$ & $0$ & $0$ \\
$\mathbf{+ 5/2}$ & $0$ & $0$ & $\frac{8 \sqrt{3} B^{4}_{4}}{77} + \frac{80 \sqrt{3} B^{4}_{6}}{1287}$ & $0$ \\
$\mathbf{+ 7/2}$ & $0$ & $0$ & $0$ & $\frac{8 \sqrt{35} B^{4}_{4}}{385} - \frac{80 \sqrt{35} B^{4}_{6}}{3003}$ \\
\toprule
\end{tabular}


\begin{tabular}{ccccc}

      & $\mathbf{+ 1/2}$ & $\mathbf{+ 3/2}$ & $\mathbf{+ 5/2}$ & $\mathbf{+ 7/2}$ \\
\hline
$\mathbf{- 7/2}$ & $\frac{8 \sqrt{35} B^{4}_{4}}{385} - \frac{80 \sqrt{35} B^{4}_{6}}{3003}$ & $0$ & $0$ & $0$ \\
$\mathbf{- 5/2}$ & $0$ & $\frac{8 \sqrt{3} B^{4}_{4}}{77} + \frac{80 \sqrt{3} B^{4}_{6}}{1287}$ & $0$ & $0$ \\
$\mathbf{- 3/2}$ & $0$ & $0$ & $\frac{8 \sqrt{3} B^{4}_{4}}{77} + \frac{80 \sqrt{3} B^{4}_{6}}{1287}$ & $0$ \\
$\mathbf{- 1/2}$ & $0$ & $0$ & $0$ & $\frac{8 \sqrt{35} B^{4}_{4}}{385} - \frac{80 \sqrt{35} B^{4}_{6}}{3003}$ \\
$\mathbf{+ 1/2}$ & $\frac{10 B^{0}_{2}}{21} + \frac{72 B^{0}_{4}}{77} + \frac{400 B^{0}_{6}}{429}$ & $0$ & $0$ & $0$ \\
$\mathbf{+ 3/2}$ & $0$ & $\frac{2 B^{0}_{2}}{7} - \frac{24 B^{0}_{4}}{77} - \frac{240 B^{0}_{6}}{143}$ & $0$ & $0$ \\
$\mathbf{+ 5/2}$ & $0$ & $0$ & $- \frac{2 B^{0}_{2}}{21} - \frac{104 B^{0}_{4}}{77} + \frac{400 B^{0}_{6}}{429}$ & $0$ \\
$\mathbf{+ 7/2}$ & $0$ & $0$ & $0$ & $- \frac{2 B^{0}_{2}}{3} + \frac{8 B^{0}_{4}}{11} - \frac{80 B^{0}_{6}}{429}$ \\
\toprule
\end{tabular}

\end{table*}

\begin{table*}[h!]
\centering
\small
\caption{Clebsch--Gordan transformation matrix ($U^\text{CG}$) between the uncoupled $\ket{m_l, m_s}$ basis (columns) and coupled total angular momentum basis $\ket{j,m_j}$ (rows) for an f$^1$ configuration ($l=3$, $s=\tfrac{1}{2}$).}
\label{tab-si:ucg}
\renewcommand{\arraystretch}{1.2}
\setlength{\tabcolsep}{1pt}
\begin{tabular}{rcccccccccccccc}
\toprule
$U^{\text{CG}}$
& $\mathbf{\ket{-3,-\frac12}}$ & $\mathbf{\ket{-2,-\frac12}}$ & $\mathbf{\ket{-1,-\frac12}}$ & $\mathbf{\ket{0,-\frac12}}$
& $\mathbf{\ket{+1,-\frac12}}$ & $\mathbf{\ket{+2,-\frac12}}$ & $\mathbf{\ket{+3,-\frac12}}$
& $\mathbf{\ket{-3,+\frac12}}$ & $\mathbf{\ket{-2,+\frac12}}$ & $\mathbf{\ket{-1,+\frac12}}$ & $\mathbf{\ket{0,+\frac12}}$
& $\mathbf{\ket{+1,+\frac12}}$ & $\mathbf{\ket{+2,+\frac12}}$ & $\mathbf{\ket{+3,+\frac12}}$ \\ \hline

$\mathbf{\bra{\tfrac52,-\tfrac52}}$
& $0$ & $\tfrac{\sqrt7}{7}$ & $0$ & $0$ & $0$ & $0$ & $0$
& $-\tfrac{\sqrt{42}}{7}$ & $0$ & $0$ & $0$ & $0$ & $0$ & $0$ \\

$\mathbf{\bra{\tfrac52,-\tfrac32}}$
& $0$ & $0$ & $\tfrac{\sqrt{14}}{7}$ & $0$ & $0$ & $0$ & $0$
& $0$ & $-\tfrac{\sqrt{35}}{7}$ & $0$ & $0$ & $0$ & $0$ & $0$ \\

$\mathbf{\bra{\tfrac52,-\tfrac12}}$
& $0$ & $0$ & $0$ & $\tfrac{\sqrt{21}}{7}$ & $0$ & $0$ & $0$
& $0$ & $0$ & $-\tfrac{2\sqrt7}{7}$ & $0$ & $0$ & $0$ & $0$ \\

$\mathbf{\bra{\tfrac52,+\tfrac12}}$
& $0$ & $0$ & $0$ & $0$ & $\tfrac{2\sqrt7}{7}$ & $0$ & $0$
& $0$ & $0$ & $0$ & $-\tfrac{\sqrt{21}}{7}$ & $0$ & $0$ & $0$ \\

$\mathbf{\bra{\tfrac52,+\tfrac32}}$
& $0$ & $0$ & $0$ & $0$ & $0$ & $\tfrac{\sqrt{35}}{7}$ & $0$
& $0$ & $0$ & $0$ & $0$ & $-\tfrac{\sqrt{14}}{7}$ & $0$ & $0$ \\

$\mathbf{\bra{\tfrac52,+\tfrac52}}$
& $0$ & $0$ & $0$ & $0$ & $0$ & $0$ & $\tfrac{\sqrt{42}}{7}$
& $0$ & $0$ & $0$ & $0$ & $0$ & $-\tfrac{\sqrt7}{7}$ & $0$ \\

$\mathbf{\bra{\tfrac72,-\tfrac72}}$
& $1$ & $0$ & $0$ & $0$ & $0$ & $0$ & $0$
& $0$ & $0$ & $0$ & $0$ & $0$ & $0$ & $0$ \\

$\mathbf{\bra{\tfrac72,-\tfrac52}}$
& $0$ & $\tfrac{\sqrt{42}}{7}$ & $0$ & $0$ & $0$ & $0$ & $0$
& $\tfrac{\sqrt7}{7}$ & $0$ & $0$ & $0$ & $0$ & $0$ & $0$ \\

$\mathbf{\bra{\tfrac72,-\tfrac32}}$
& $0$ & $0$ & $\tfrac{\sqrt{35}}{7}$ & $0$ & $0$ & $0$ & $0$
& $0$ & $\tfrac{\sqrt{14}}{7}$ & $0$ & $0$ & $0$ & $0$ & $0$ \\

$\mathbf{\bra{\tfrac72,-\tfrac12}}$
& $0$ & $0$ & $0$ & $\tfrac{2\sqrt7}{7}$ & $0$ & $0$ & $0$
& $0$ & $0$ & $\tfrac{\sqrt{21}}{7}$ & $0$ & $0$ & $0$ & $0$ \\

$\mathbf{\bra{\tfrac72,+\tfrac12}}$
& $0$ & $0$ & $0$ & $0$ & $\tfrac{\sqrt{21}}{7}$ & $0$ & $0$
& $0$ & $0$ & $0$ & $\tfrac{2\sqrt7}{7}$ & $0$ & $0$ & $0$ \\

$\mathbf{\bra{\tfrac72,+\tfrac32}}$
& $0$ & $0$ & $0$ & $0$ & $0$ & $\tfrac{\sqrt{14}}{7}$ & $0$
& $0$ & $0$ & $0$ & $0$ & $\tfrac{\sqrt{35}}{7}$ & $0$ & $0$ \\

$\mathbf{\bra{\tfrac72,+\tfrac52}}$
& $0$ & $0$ & $0$ & $0$ & $0$ & $0$ & $\tfrac{\sqrt7}{7}$
& $0$ & $0$ & $0$ & $0$ & $0$ & $\tfrac{\sqrt{42}}{7}$ & $0$ \\

$\mathbf{\bra{\tfrac72,+\tfrac72}}$
& $0$ & $0$ & $0$ & $0$ & $0$ & $0$ & $0$
& $0$ & $0$ & $0$ & $0$ & $0$ & $0$ & $1$ \\
\toprule
\end{tabular}
\end{table*}


\begin{table*}[h!]
\centering
\caption{The off-diagonal block of the CF model matrix showing expressions for the coupling elements between $j=5/2$ and $j^\prime=7/2$ manifolds. This is $V_{j,j^\prime}^\text{CF}$ in Equation \ref{eq:vcf-j}.}
\label{tab-si:cf-jmixing-placeholders}
\renewcommand{\arraystretch}{1.5}
\begin{tabular}{l}
\toprule
$\mathbf{A}$: $39 B^{4}_{4} - 50 B^{4}_{6}$  \\
$\mathbf{B}$: $143 B^{0}_{2} - 520 B^{0}_{4} + 280 B^{0}_{6}$ \\
$\mathbf{C}$: $78 B^{4}_{4} + 175 B^{4}_{6}$ \\
$\mathbf{D}$: $429 B^{0}_{2} + 2080 B^{0}_{4} - 4200 B^{0}_{6}$ \\
$\mathbf{E}$: $13 B^{4}_{4} - 35 B^{4}_{6}$ \\
$\mathbf{F}$: $143 B^{0}_{2} + 1300 B^{0}_{4} + 7000 B^{0}_{6}$ \\
\toprule
\end{tabular}
\end{table*}


\begin{table*}[h!]
\centering
\caption{\emph{Ab initio} crystal-field matrix elements (cm$^{-1}$ units) within the $j=5/2$ manifold, derived from a SOCI calculation on a Ce$^{3+}$ ion in $D_\text{4h}$ symmetry; it should be brought in correspondence with the model CF matrix of Table \ref{tab-si:cf-j52}.}
\label{tab-si:hcf-j52}
\renewcommand{\arraystretch}{1.2}
    \begin{tabular}{cccccc}
    \toprule
    $-235.4-0.0i$ & $0i$ & $0i$ & $0i$ & $108.3+0.0i$ & $0i$ \\
    $0i$ & $-1637.7-0.0i$ & $0i$ & $0i$ & $0i$ & $108.3+0.0i$ \\
    $0i$ & $0i$ & $-2204.4+0.0i$ & $0i$ & $0i$ & $0i$ \\
    $0i$ & $0i$ & $0i$ & $-2204.4-0.0i$ & $0i$ & $0i$ \\
    $108.3-0.0i$ & $0i$ & $0i$ & $0i$ & $-1637.7+0.0i$ & $0i$ \\
    $0i$ & $108.3-0.0i$ & $0i$ & $0i$ & $0i$ & $-235.4+0.0i$ \\
    \toprule
    \end{tabular}
\end{table*}


\begin{table*}[h!]
\centering
\caption{\emph{Ab initio} crystal-field matrix elements (cm$^{-1}$ units) within the $j^\prime=7/2$ manifold, derived from a SOCI calculation on a Ce$^{3+}$ ion in $D_\text{4h}$ symmetry; it should be brought in correspondence with the model CF matrix of Table \ref{tab-si:cf-j72}.}
\label{tab-si:hcf-j72}
\renewcommand{\arraystretch}{1.2}
\setlength{\tabcolsep}{0.5pt}
  \begin{tabular}{cccccccc}
    \toprule
    $2363.1-0.0i$ & $0i$ & $0i$ & $0i$ & $87.3-0.0i$ & $0i$ & $0i$ & $0i$ \\
    $0i$ & $1127.1-0.0i$ & $0i$ & $0i$ & $0i$ & $111.9-0.0i$ & $0i$ & $0i$ \\
    $0i$ & $0i$ & $451.1-0.0i$ & $0i$ & $0i$ & $0i$ & $111.9-0.0i$ & $0i$ \\
    $0i$ & $0i$ & $0i$ & $136.3+0.0i$ & $0i$ & $0i$ & $0i$ & $87.3+0.0i$ \\
    $87.3+0.0i$ & $0i$ & $0i$ & $0i$ & $136.3+0.0i$ & $0i$ & $0i$ & $0i$ \\
    $0i$ & $111.9+0.0i$ & $0i$ & $0i$ & $0i$ & $451.1-0.0i$ & $0i$ & $0i$ \\
    $0i$ & $0i$ & $111.9+0.0i$ & $0i$ & $0i$ & $0i$ & $1127.1+0.0i$ & $0i$ \\
    $0i$ & $0i$ & $0i$ & $87.3-0.0i$ & $0i$ & $0i$ & $0i$ & $2363.1+0.0i$ \\
    \toprule
  \end{tabular}
\end{table*}


\begin{table*}[h!]
    \small
    \centering
    \caption{\emph{Ab initio} CF coupling elements ($\icm$ units) between the $j=5/2$ and $j^\prime=7/2$ multiplets, derived from a SOCI calculation on a Ce$^{3+}$ ion in a field of point-charges with $O_\text{h}$ symmetry.$^{a}$}
    \label{tab-si:hcf-jmixing-oh}
    \renewcommand{\arraystretch}{1.3}
    \setlength{\tabcolsep}{3pt}
  \begin{tabular}{ccccccc}
    \toprule
  $\hat{H}\mathrm{(CSH)}$   & $\ket{\frac{5}{2},-\frac{5}{2}}$ & $\ket{\frac{5}{2},-\frac{3}{2}}$ & $\ket{\frac{5}{2},-\frac{1}{2}}$ & $\ket{\frac{5}{2},+\frac{1}{2}}$ & $\ket{\frac{5}{2},+\frac{3}{2}}$ & $\ket{\frac{5}{2},+\frac{5}{2}}$ \\ \hline
$\bra{\frac{7}{2},-\frac{7}{2}}$ & $0i$ & $0i$ & $0i$ & $100.1 - 0.0i$ & $0i$ & $0i$ \\
$\bra{\frac{7}{2},-\frac{5}{2}}$ & $-107.9 - 0.0i$ & $0i$ & $0i$ & $0i$ & $82.4 + 0.0i$ & $0i$ \\
$\bra{\frac{7}{2},-\frac{3}{2}}$ & $0i$ & $108.9 - 0.0i$ & $0i$ & $0i$ & $0i$ & $73.3 - 0.0i$ \\
$\bra{\frac{7}{2},-\frac{1}{2}}$ & $0i$ & $0i$ & $83.6 + 0.0i$ & $0i$ & $0i$ & $0i$ \\
$\bra{\frac{7}{2},+\frac{1}{2}}$ & $0i$ & $0i$ & $0i$ & $-83.6 + 0.0i$ & $0i$ & $0i$ \\
$\bra{\frac{7}{2},+\frac{3}{2}}$ & $-73.3 - 0.0i$ & $0i$ & $0i$ & $0i$ & $-108.9 - 0.0i$ & $0i$ \\
$\bra{\frac{7}{2},+\frac{5}{2}}$ & $0i$ & $-82.4 - 0.0i$ & $0i$ & $0i$ & $0i$ & $107.9 - 0.0i$ \\
$\bra{\frac{7}{2},+\frac{7}{2}}$ & $0i$ & $0i$ & $-100.1 + 0.0i$ & $0i$ & $0i$ & $0i$ \\
    \toprule
  \end{tabular}
\flushleft{$^a$For comparison, the energy separation between the $j=\tfrac{5}{2}$ and $j^\prime =\tfrac{7}{2}$ manifolds is $\tfrac{7}{2}\zeta_\text{4f}=2375$ cm$^{-1}$.}
\end{table*}

\begin{table*}[h!]
    \small
    \centering
    \caption{\emph{Ab initio} CF coupling elements ($\icm$ units) between the $j=5/2$ and $j^\prime=7/2$ multiplets, derived from a SOCI calculation on a Ce$^{3+}$ ion in a field of point-charges with $T_\text{d}$ symmetry.$^{a}$}
    \label{tab-si:hcf-jmixing-td}
    \renewcommand{\arraystretch}{1.3}
    \setlength{\tabcolsep}{3pt}
  \begin{tabular}{ccccccc}
    \toprule
  $\hat{H}\mathrm{(CSH)}$   & $\ket{\frac{5}{2},-\frac{5}{2}}$ & $\ket{\frac{5}{2},-\frac{3}{2}}$ & $\ket{\frac{5}{2},-\frac{1}{2}}$ & $\ket{\frac{5}{2},+\frac{1}{2}}$ & $\ket{\frac{5}{2},+\frac{3}{2}}$ & $\ket{\frac{5}{2},+\frac{5}{2}}$ \\ \hline
$\bra{\frac{7}{2},-\frac{7}{2}}$ & $0i$ & $0i$ & $0i$ & $-43.0 - 0.0i$ & $0i$ & $0i$ \\
$\bra{\frac{7}{2},-\frac{5}{2}}$ & $51.5 - 0.0i$ & $0i$ & $0i$ & $0i$ & $-47.1 - 0.0i$ & $0i$ \\
$\bra{\frac{7}{2},-\frac{3}{2}}$ & $0i$ & $-53.4 - 0.0i$ & $0i$ & $0i$ & $0i$ & $-28.0 - 0.0i$ \\
$\bra{\frac{7}{2},-\frac{1}{2}}$ & $0i$ & $0i$ & $-35.8 + 0.0i$ & $0i$ & $0i$ & $0i$ \\
$\bra{\frac{7}{2},+\frac{1}{2}}$ & $0i$ & $0i$ & $0i$ & $35.8 - 0.0i$ & $0i$ & $0i$ \\
$\bra{\frac{7}{2},+\frac{3}{2}}$ & $28.0 + 0.0i$ & $0i$ & $0i$ & $0i$ & $53.4 - 0.0i$ & $0i$ \\
$\bra{\frac{7}{2},+\frac{5}{2}}$ & $0i$ & $47.1 + 0.0i$ & $0i$ & $0i$ & $0i$ & $-51.5 + 0.0i$ \\
$\bra{\frac{7}{2},+\frac{7}{2}}$ & $0i$ & $0i$ & $43.0 - 0.0i$ & $0i$ & $0i$ & $0i$ \\
    \toprule
  \end{tabular}
\flushleft{$^a$For comparison, the energy separation between the $j=\tfrac{5}{2}$ and $j^\prime =\tfrac{7}{2}$ manifolds is $\tfrac{7}{2}\zeta_\text{4f}=2374$ cm$^{-1}$.}
\end{table*}

\begin{table*}[h!]
    \small
    \centering
    \caption{\emph{Ab initio} CF coupling elements ($\icm$ units) between the $j=5/2$ and $j^\prime=7/2$ multiplets, derived from a SOCI calculation on a Ce$^{3+}$ ion in a field of point-charges with $C_\text{2v}$ symmetry.$^{a}$}
    \label{tab-si:hcf-jmixing-c2v}
    \renewcommand{\arraystretch}{1.3}
    \setlength{\tabcolsep}{3pt}
  \begin{tabular}{ccccccc}
    \toprule
  $\hat{H}\mathrm{(CSH)}$   & $\ket{\frac{5}{2},-\frac{5}{2}}$ & $\ket{\frac{5}{2},-\frac{3}{2}}$ & $\ket{\frac{5}{2},-\frac{1}{2}}$ & $\ket{\frac{5}{2},+\frac{1}{2}}$ & $\ket{\frac{5}{2},+\frac{3}{2}}$ & $\ket{\frac{5}{2},+\frac{5}{2}}$ \\ \hline
$\bra{\frac{7}{2},-\frac{7}{2}}$ & $0i$ & $-161.7 + 0.0i$ & $0i$ & $16.9 + 0.0i$ & $0i$ & $-3.0 + 0.0i$ \\
$\bra{\frac{7}{2},-\frac{5}{2}}$ & $-21.9 + 0.0i$ & $0i$ & $-182.9 + 0.0i$ & $0i$ & $20.1 + 0.0i$ & $0i$ \\
$\bra{\frac{7}{2},-\frac{3}{2}}$ & $0i$ & $-31.1 + 0.0i$ & $0i$ & $-154.8 + 0.0i$ & $0i$ & $10.6 + 0.0i$ \\
$\bra{\frac{7}{2},-\frac{1}{2}}$ & $53.9 + 0.0i$ & $0i$ & $-11.7 + 0.0i$ & $0i$ & $-109.8 + 0.0i$ & $0i$ \\
$\bra{\frac{7}{2},+\frac{1}{2}}$ & $0i$ & $109.8 + 0.0i$ & $0i$ & $11.7 + 0.0i$ & $0i$ & $-53.9 + 0.0i$ \\
$\bra{\frac{7}{2},+\frac{3}{2}}$ & $-10.6 + 0.0i$ & $0i$ & $154.8 + 0.0i$ & $0i$ & $31.1 + 0.0i$ & $0i$ \\
$\bra{\frac{7}{2},+\frac{5}{2}}$ & $0i$ & $-20.1 + 0.0i$ & $0i$ & $182.9 + 0.0i$ & $0i$ & $21.9 + 0.0i$ \\
$\bra{\frac{7}{2},+\frac{7}{2}}$ & $3.0 + 0.0i$ & $0i$ & $-16.9 + 0.0i$ & $0i$ & $161.7 + 0.0i$ & $0i$ \\
    \toprule
  \end{tabular}
\flushleft{$^a$For comparison, the energy separation between the $j=\tfrac{5}{2}$ and $j^\prime =\tfrac{7}{2}$ manifolds is $\tfrac{7}{2}\zeta_\text{4f}=2374$ cm$^{-1}$.}
\end{table*}

\end{document}